\PassOptionsToPackage{unicode}{hyperref}
\PassOptionsToPackage{hyphens}{url}
\PassOptionsToPackage{dvipsnames,svgnames,x11names}{xcolor}
\documentclass[
  11pt,
]{article}
\usepackage{xcolor}
\usepackage[margin=1in]{geometry}
\usepackage{amsmath,amssymb}
\usepackage{iftex}
\ifPDFTeX
  \usepackage[T1]{fontenc}
  \usepackage[utf8]{inputenc}
  \usepackage{textcomp} % provide euro and other symbols
\else % if luatex or xetex
  \usepackage{unicode-math} % this also loads fontspec
  \defaultfontfeatures{Scale=MatchLowercase}
  \defaultfontfeatures[\rmfamily]{Ligatures=TeX,Scale=1}
\fi
\usepackage{lmodern}
\ifPDFTeX\else
\fi
\IfFileExists{upquote.sty}{\usepackage{upquote}}{}
\IfFileExists{microtype.sty}{% use microtype if available
  \usepackage[]{microtype}
  \UseMicrotypeSet[protrusion]{basicmath} % disable protrusion for tt fonts
}{}
\makeatletter
\@ifundefined{KOMAClassName}{% if non-KOMA class
  \IfFileExists{parskip.sty}{%
    \usepackage{parskip}
  }{% else
    \setlength{\parindent}{0pt}
    \setlength{\parskip}{6pt plus 2pt minus 1pt}}
}{% if KOMA class
  \KOMAoptions{parskip=half}}
\makeatother
\usepackage{longtable,booktabs,array}
\usepackage{caption}
\usepackage{calc} % for calculating minipage widths
\usepackage{etoolbox}
\makeatletter
\patchcmd\longtable{\par}{\if@noskipsec\mbox{}\fi\par}{}{}
\makeatother
\IfFileExists{footnotehyper.sty}{\usepackage{footnotehyper}}{\usepackage{footnote}}
\makesavenoteenv{longtable}
\usepackage{graphicx}
\makeatletter
\newsavebox\pandoc@box
\newcommand*\pandocbounded[1]{% scales image to fit in text height/width
  \sbox\pandoc@box{#1}%
  \Gscale@div\@tempa{\textheight}{\dimexpr\ht\pandoc@box+\dp\pandoc@box\relax}%
  \Gscale@div\@tempb{\linewidth}{\wd\pandoc@box}%
  \ifdim\@tempb\p@<\@tempa\p@\let\@tempa\@tempb\fi% select the smaller of both
  \ifdim\@tempa\p@<\p@\scalebox{\@tempa}{\usebox\pandoc@box}%
  \else\usebox{\pandoc@box}%
  \fi%
}
\def\fps@figure{htbp}
\makeatother
\usepackage{bookmark}
\IfFileExists{xurl.sty}{\usepackage{xurl}}{} % add URL line breaks if available
\hypersetup{
  colorlinks=true,
  linkcolor={Maroon},
  filecolor={Maroon},
  citecolor={Blue},
  urlcolor={Blue},
  pdfcreator={LaTeX via pandoc}}

\author{Mingguang Chen\textsuperscript{1,$*$}\\[4pt]
{\small \textsuperscript{1}DeepGrounding}\\[2pt]
{\small \textsuperscript{$*$}Corresponding author. Email: \href{mailto:deepgroundingai@gmail.com}{deepgroundingai@gmail.com}}}
\date{August 2026}

\begin{document}

\section{Title}\label{title}

Fermi-Level Metal-d Character of Group-6 Bis-Hexahapto Bilayer Graphene

Mingguang Chen1,*

1 DeepGrounding

*Corresponding author. Email: deepgroundingai@gmail.com

\section{Introduction}\label{introduction}

Superconductivity in graphite intercalation compounds is a settled
experimental fact and a well-understood one. CaC6 superconducts at 11.5
K and YbC6 at 6.5 K {[}1{]}, and density functional theory established
early that the pairing in CaC6 is phonon mediated with a contribution
from the intercalant Fermi surface {[}2{]}. The decisive
electronic-structure ingredient, identified by Csányi and co-workers
{[}3{]}, is the interlayer band: a nearly free-electron state residing
in the gallery between the carbon sheets rather than on the atoms.
Across the intercalated graphites, superconductivity tracks whether that
band is occupied, and its occupancy is supplied by charge donated from
an electropositive intercalant. Compounds in which the donated electrons
instead fill the graphene pi* states do not superconduct. The same logic
carries to the two-dimensional limit: Li decoration of monolayer
graphene was predicted {[}4{]} and subsequently observed {[}5{]} to
support superconductivity at a few kelvin, and Ca-intercalated bilayer
graphene superconducts at 4 K {[}6{]} whereas the Li-intercalated
bilayer analogue on the same substrate does not {[}6{]}.

A quite different way of binding a metal to a graphitic surface has been
developed in parallel. In bis-hexahapto (eta6-eta6) coordination, a
transition metal binds covalently to the pi systems of two carbon sheets
at once, forming a sandwich analogous to bis(benzene)chromium. Applied
to carbon nanotubes {[}7,8{]} and to graphene nanoplatelets {[}9{]},
this chemistry interconnects adjacent carbon surfaces and raises bulk
conductivity while, unlike sp3 functionalisation, leaving the sheets'
conjugation substantially intact {[}9{]}. The defining feature of the
interaction, and the reason it is chemically attractive, is that it is
covalent: metal d and carbon pi orbitals hybridise directly, with little
net charge transfer to the sheets. Indeed, the absence of charge
transfer is exactly why transition metals cannot be intercalated into
graphite by the routes that work for alkali and alkaline-earth metals.

It is natural to ask whether such covalently intercalated sandwiches
might also superconduct, particularly if the metal atoms were ordered
into a periodic sublattice, and this has been suggested for
bis-hexahapto intercalated carbon systems {[}10{]}. The suggestion
deserves a careful answer, because the two mechanisms above pull in
opposite directions. The property that makes graphite intercalation
compounds superconduct is charge donation into an interlayer band; the
property that defines bis-hexahapto coordination is the absence of that
donation. Stated this way, the question is sharp and answerable from
first principles: do the states at the Fermi level in a transition-metal
bis-hexahapto sandwich retain the carbon-derived, interstitial-like
character associated with an occupied interlayer band in ionic graphite
intercalation compounds? All candidate systems here are metallic on
dense Brillouin-zone meshes; the operative distinction is orbital
composition, not the presence or absence of a Fermi surface. We do not
compute electron-phonon coupling or a critical temperature.

Existing calculations bear on this only partially, and only for
chromium. Embedding of transition-metal atoms in graphene already
produces a rich structure-magnetism landscape that depends sensitively
on local coordination {[}11{]}. Lucht and co-workers {[}12{]} surveyed
the first-row transition metals in a honeycomb sublattice between
AA-stacked sheets and found that for the earlier metals, including Cr,
the Dirac cone ``loses its linearity and becomes quadratic and even
gapped''; chromium carries a moment of about 5 muB and orders
antiferromagnetically, and the two elements the authors identify as
viable candidates for their proposed magnetic-exchange route are
vanadium and nickel, not chromium. Separately, Li and co-workers
{[}13{]} constructed Cr-intercalated bilayer graphene from
bis(benzene)chromium building blocks and reported that d-pi interaction
drives each C6CrC6 unit to an 18-electron closed shell, opening a gap.
Neither study sweeps intercalant coverage continuously, neither treats
molybdenum or tungsten, neither tests the interlayer-band criterion
directly, and neither addresses whether an ordered metal sublattice is
thermodynamically attainable in the first place -- a pointed question
for chromium, which is mobile on graphitic surfaces and a well-known
cluster former.

Here we address those gaps for the group-6 series Cr, Mo and W -- the
same series whose hexacarbonyl reagents span the experimentally observed
reactivity ordering Cr \textgreater{} Mo \textgreater{} W on graphene
and graphene-nanoplatelet films {[}9{]} (detailed further in the present
author's dissertation {[}10{]}) -- at the C12M and C6M coverages that
bracket isolated sandwich units and a metal-metal bonded honeycomb, with
additional C24M and C48M cells where magnetism requires them, and in
both AA and AB stacking. We begin from a geometric observation that
appears not to have been stated: eta6-eta6 coordination requires a
hexagon centre in both sheets at the same in-plane position, which AA
stacking provides exactly and AB stacking forbids exactly. The
bis-hexahapto motif therefore carries an intrinsic stacking penalty
against the ground-state stacking of bilayer graphene. We then
determine, with the same protocol applied to bulk CaC6 and to Ca- and
Li-intercalated bilayers as controls, whether the group-6 sandwiches are
thermodynamically stable against clustering, what their magnetic ground
states are, and -- the central question -- whether the states at the
Fermi level in these covalent bilayers retain the carbon-derived,
interstitial-like character associated with an occupied interlayer band
in the ionic references.

\section{Computational methods}\label{computational-methods}

\section{Electronic structure}\label{electronic-structure}

All calculations use density functional theory as implemented in GPAW
25.7.0 {[}14{]} with the projector augmented-wave method {[}15{]},
driven through ASE 3.29.0 {[}16{]}. Exchange and correlation are treated
with the PBE generalised gradient approximation {[}17{]}. Long-range
dispersion, which controls the interlayer binding of graphene sheets and
is absent from PBE, is added as Grimme's D3 correction with
Becke-Johnson damping {[}18,19{]} via \texttt{simple-dftd3} 1.4.0,
applied as an additive correction to the PBE energy and forces.

\emph{Basis.} Wavefunctions are expanded in plane waves with a 450 eV
cutoff. This is a deliberate departure from the
linear-combination-of-atomic-orbitals (LCAO) approach that would be the
cheaper option for cells of this size. In LCAO with a
double-zeta-polarised numerical basis, the basis functions on the
intercalated metal and on the twelve carbon atoms coordinated to it
become linearly dependent at the equilibrium gallery spacing: the
overlap matrix loses positive-definiteness and the generalised
eigenvalue problem cannot be solved (GPAW raises a Cholesky failure
during LCAO diagonalisation of C12Cr). This is a numerical breakdown
rather than a loss of accuracy, and it is not repaired by any choice of
mixing or convergence criteria. Plane waves have no such overlap
pathology, and for the 13-49-atom cells used here they cost
approximately the same per self-consistency step.

\emph{Brillouin-zone sampling.} Monkhorst-Pack meshes are used
throughout, with in-plane density held approximately constant across
cell sizes: 9x9x1 for the sqrt(3)xsqrt(3) cells, 6x6x1 and 5x5x1 for the
2x1 and 2x2 supercells, and 12x12x12 for bulk metal references. Bulk
CaC6 uses 9x9x3. Occupations are smeared with a Fermi-Dirac distribution
of width 0.05 eV. This width is smaller than the value conventionally
used for metals because the central quantity here is whether a small gap
opens: a 0.1 eV smearing measurably erodes a gap of a few tenths of an
eV and can therefore manufacture apparent metallicity. It is nonetheless
large enough to keep the self-consistency cycle stable in the
spin-polarised calculations, which sit close to a magnetic instability.
The insensitivity of the reported gaps to this choice is documented in
the Supporting Information.

\emph{Spin.} Spin-polarised calculations use separate mixing of the
total and magnetisation densities (GPAW's \texttt{MixerDif}, beta =
0.02, eight-step history). With the default common mixing, the C12Cr
self-consistency cycle drives the energy and density to convergence
within roughly twenty iterations but leaves the eigenstate residual near
10\^{}-2 for more than forty, with the magnetic moment oscillating; this
is the standard failure mode for a system near a magnetic instability
and is resolved by treating the two densities separately. Non-magnetic
and ferromagnetic solutions are compared for the C12 cells.
Antiferromagnetic order is additionally tested where the simulation cell
contains enough metal sites to represent it. The triangular one-metal
C12 cell cannot represent an antiferromagnetic pattern, so no such
result is inferred from that cell. C12W adopts a weak ferromagnetic
ground state (0.65 muB; see Results §6); C12Cr and C12Mo are
non-magnetic. Each compared magnetic state is independently relaxed to
the same numerical criteria; constrained-moment scans, where used, hold
geometry fixed. The full calculated energy spread is reported rather
than only a winning label.

\emph{Constrained-moment scans.} Unconstrained spin-polarised
calculations are not sufficient here, and relying on them would be
unsound. For C12Cr the unconstrained cycle ran past 150 iterations
without converging, with the total moment oscillating between 0.19 and
0.92 muB and the density residual stalled near 10\^{}-3 -- the behaviour
of a calculation hunting between nearly degenerate magnetic and
non-magnetic basins rather than one converging slowly. Magnetic ground
states are therefore established from scans of the total energy at fixed
total magnetic moment, E(M), with the geometry held fixed across each
scan so that magnetic and structural energies are not confounded. Each
constrained calculation has a single well-defined solution and converges
normally. This is the standard treatment for itinerant magnets near a
magnetic instability, and it is strictly more informative than a single
unconstrained number: it reveals the depth of the magnetic minimum, the
presence or absence of competing minima, and whether a high-moment
solution reported elsewhere exists on this energy surface at all.

A practical consequence worth stating: a single unconstrained
spin-polarised moment for these materials should not be trusted without
evidence that the self-consistency cycle converged.

No Hubbard correction is used in the reported production data. A U
parameter for itinerant metal--graphene hybrid states would require an
independent calibration; presenting an arbitrary U sweep as uncertainty
would instead mix different Hamiltonians without assigning any of them
evidential priority. A fixed-geometry DFT+U spot check on production
C12Cr (U = 0, 2, 4, 6 eV; \texttt{dft/dftu\_spotcheck.json}) is recorded
in the SI to confirm that moderate U does not reopen a gap or overturn
the non-magnetic metallic ground state used in the production
comparison.

\section{Structures}\label{structures}

Bilayer graphene is built in the sqrt(3)xsqrt(3)R30 cell, which contains
six carbon atoms and three hexagon centres per layer. Metal atoms are
placed at hexagon centres in the gallery, giving the coverage series C6M
(two of three sites, a honeycomb metal sublattice), C12M (one of three,
triangular), and C24M and C48M in 2x1 and 2x2 supercells of the same
cell.

The two stackings are treated on a different footing, for a geometric
reason that is central to this work. In AA stacking the hexagon centres
of the two layers are collinear, so a metal atom can be eta-6
coordinated to both sheets simultaneously; we verify numerically that
the projected offset between the two layers' hexagon centres is zero and
that the metal has twelve equidistant carbon neighbours. In AB (Bernal)
stacking, a carbon atom of the opposing layer lies exactly over every
hexagon centre, so bis-hexahapto coordination is forbidden by symmetry
rather than merely disfavoured. AB structures therefore serve as
controls that price the stacking penalty, not as competing bis-hexahapto
candidates.

Initial gallery separations are taken from measured structures rather
than estimated: 3.20 A for Cr, from the 2.14 A Cr-C distance in
bis(benzene)chromium; 3.6 A for Mo and W; and the experimental
graphite-intercalation-compound spacings of 4.524 A for Ca and 3.706 A
for Li.

Geometries are relaxed with the BFGS algorithm. A coarse k mesh is used
only to generate the starting trajectory; every structure entering a
quantitative comparison is then refined with the production k mesh to a
maximum residual force of 0.02 eV/A. In-plane lattice constants are
relaxed separately from the atomic positions, with the vacuum direction
held fixed. Records that fail the production-mesh force test are
excluded from headline comparisons until this refinement is complete.

\section{Reference states and derived
quantities}\label{reference-states-and-derived-quantities}

Formation energies per intercalated metal atom are referenced two
independent ways: against the bulk metal, which asks whether the ordered
lattice is stable against segregation into metal, and against isolated
metal atoms, which is the appropriate reference for the metal-vapour
deposition route used experimentally {[}9,10{]}. Bulk chromium is
computed in the conventional two-atom cell so that its antiferromagnetic
ground state is representable. The two references are required to rank
structures identically; any discrepancy is reported.

\emph{Interlayer-band character.} The criterion of Csányi and co-workers
ties graphite-intercalation-compound superconductivity to occupancy of a
nearly free-electron interlayer band whose states are carbon-derived and
spatially delocalised in the gallery. We test a necessary structural
ingredient of that picture -- whether gallery density at E\_F survives
masking of atomic spheres and projects onto carbon p rather than metal d
-- not electron-phonon coupling or a superconducting transition. For
every state within 1.5 eV of the Fermi level we integrate the
pseudo-wavefunction density in the central 45\% of the carbon gallery.
We record both the raw gallery integral, G\_i, and the integral I\_i
remaining after masking a 1.30 A sphere around every nucleus. The
primary observable is a same-state ratio of weighted sums, sum\_i(q\_i
I\_i)/sum\_i(q\_i G\_i), where q\_i is the irreducible-k-point weight
multiplied by a Gaussian Fermi-level kernel of sigma = 0.10 eV. This
avoids the legacy ratio of independent maxima, whose numerator and
denominator can come from different states. The absolute weighted
interstitial integral is reported separately. Bulk CaC6 is used as a
construct-positive benchmark, and the relevant state is checked by
direct band-resolved density rather than a scalar score alone. Mesh,
kernel-width, mask-radius and gallery-width sensitivities follow the
prospectively frozen specification in \texttt{ANALYSIS\_SPEC.md}.

\emph{Orbital projection.} Element- and l-resolved projected densities
of state are integrated in a fixed window of +/- 0.15 eV about E\_F and
normalised by element, summing for each element the bound s, p and d
projector channels its PAW dataset provides. Normalising by element
rather than by orbital channel is forced by how PAW projections work:
GPAW builds a projected density of states from bound projector functions
only, and in the default datasets the calcium d projector is unbound
while lithium has no d projector at all. A Ca-d or Li-d projected
density of states is therefore identically zero across the whole energy
range as a property of the basis, not of the material, whereas Cr, Mo
and W have bound d channels that report real weight. Comparing metal d
against carbon p would set a channel that cannot be nonzero against one
that can. The residual limitation is stated rather than hidden: a
bound-projector scheme cannot register Ca-d hybridisation with the
interlayer band, so the metal fraction reported for the ionic controls
is a lower bound, and is labelled as such wherever it appears.

\section{Validation}\label{validation}

Bulk CaC6, the reported C12Cr electronic structure, and the reported
C6Cr magnetic state are treated as explicit external benchmarks rather
than assumed truths. Agreement and disagreement are both reported. In
particular, failure of a literature-reproduction check is not relabelled
as validation: it either blocks the dependent claim or is presented as a
documented discrepancy supported by independent numerical controls.
Convergence with respect to plane-wave cutoff, k-point density, smearing
width, vacuum thickness and PAW valence is documented in the Supporting
Information.

\section{Data and code
availability}\label{data-and-code-availability}

All structure builders, calculators, job drivers and analysis scripts,
together with every machine-readable record the reported numbers are
rebuilt from, are archived at DOI 10.5281/zenodo.22586543. Every
calculation writes a JSON record, and \texttt{tmbg\_collect.py}
regenerates the derived values from those records. Analysis records
include schema versions, complete numerical parameters, source-file
SHA-256 digests and software versions. The archive also contains the
analysis specification frozen before the version-2 results were
generated (\texttt{ANALYSIS\_SPEC.md}), the dated claim-to-evidence
audit (\texttt{CLAIM\_AUDIT.md}), the machine-readable integrity report,
and a manifest of SHA-256 digests for every file it contains.
Wavefunction files are not archived; they are regenerable from the
recorded parameters.

\section{Results}\label{results}

\begin{figure}
\centering
\pandocbounded{\includegraphics[keepaspectratio,alt={Figure 1: Bis-hexahapto coordination requires AA stacking. In AA (a,c) hexagon centres of both sheets are collinear, so a metal atom is eta-6 to both layers at once; in AB (b,d) a carbon of the opposing layer sits on every hexagon centre and blocks eta-6--eta-6 coordination by symmetry. Coverage sites indicate the C6M / C12M ladder.}]{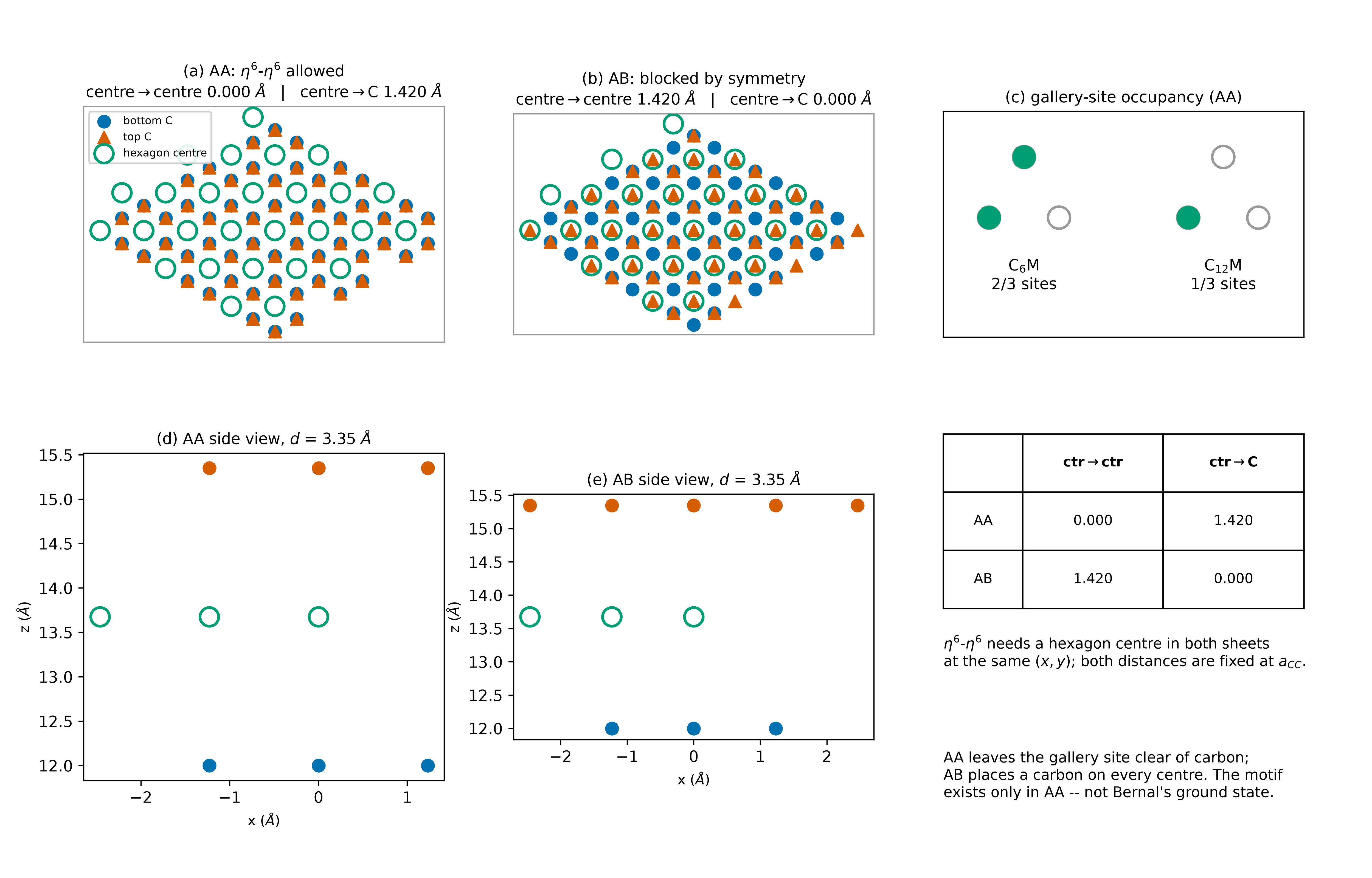}}
\caption*{Figure 1: Bis-hexahapto coordination requires AA stacking. In
AA (a,c) hexagon centres of both sheets are collinear, so a metal atom
is eta-6 to both layers at once; in AB (b,d) a carbon of the opposing
layer sits on every hexagon centre and blocks eta-6--eta-6 coordination
by symmetry. Coverage sites indicate the C6M / C12M ladder.}
\end{figure}

\section{1. Bis-hexahapto coordination requires AA
stacking}\label{bis-hexahapto-coordination-requires-aa-stacking}

The eta-6/eta-6 motif requires a hexagon centre in both sheets at the
same in-plane position. That condition is met exactly in AA stacking and
violated exactly in AB: the centre-to-centre and
centre-to-opposing-carbon distances interchange between 0 and a\_CC =
1.420 A (Figure 1). In AA the gallery site is maximally clear of carbon;
in AB a carbon atom lies directly on it. There is no intermediate case.
Bis-hexahapto coordination is therefore forbidden by symmetry in
Bernal-stacked bilayer graphene, rather than merely disfavoured -- a
stronger statement than a binding-energy comparison, and one that holds
independently of the electronic-structure method.

In the relaxed AA structures the metal is eta-6 to both sheets
simultaneously, with twelve equidistant carbon neighbours.

\subsection{The penalty is real, small, and electronic in
origin}\label{the-penalty-is-real-small-and-electronic-in-origin}

Relaxing the pristine bilayer in both stackings gives AB lower by 55.9
meV per sqrt(3) cell (4.66 meV per carbon), with interlayer separations
of 3.417 A (AB) and 3.768 A (AA). The sign and magnitude are as expected
for bilayer graphene, whose observed ground state is Bernal. Both
pristine geometries were re-relaxed on the same production force
criterion used for the intercalated cells
(\texttt{bilayer\_Cr\_*\_NM\_tight}); energies and the 55.9 meV stacking
penalty are unchanged to the reported precision, so the intercalated
comparisons below rest on a common numerical standard.

Evaluating the D3 dispersion term alone on the same geometries gives an
AA-AB difference of only 0.03 meV per carbon: dispersion is essentially
stacking blind here, because the pairwise carbon-carbon distance
distribution barely differs between the two stackings at fixed
separation. The stacking preference is therefore an electronic effect --
Pauli repulsion between eclipsed pi clouds in AA -- and consequently
robust to the choice of dispersion correction, which matters for a
conclusion resting on a small energy difference.

\subsection{Intercalation pays the penalty
easily}\label{intercalation-pays-the-penalty-easily}

Referenced to isolated metal atoms, the appropriate reference for the
metal-vapour deposition route used experimentally, chromium binds in the
gallery by -1.50 eV per Cr in C12Cr and -1.65 eV per Cr in C6Cr
(relative to the AB bilayer ground state). The 55.9 meV stacking penalty
is 3.7\% of that binding energy.

The geometric constraint is therefore not prohibitive. Its consequence
is positive and testable: intercalation imposes AA registry on the
intercalated region, which is distinguishable from Bernal stacking by
electron diffraction or by STEM imaging, since AA eclipses every carbon
column while AB eclipses only half.

\subsection{Intercalation reverses the stacking preference, for every
group-6
metal}\label{intercalation-reverses-the-stacking-preference-for-every-group-6-metal}

The direct test is to relax the same intercalated cell in both
stackings:

\begin{longtable}[]{@{}
  >{\raggedright\arraybackslash}p{(\linewidth - 10\tabcolsep) * \real{0.1667}}
  >{\raggedright\arraybackslash}p{(\linewidth - 10\tabcolsep) * \real{0.1667}}
  >{\raggedright\arraybackslash}p{(\linewidth - 10\tabcolsep) * \real{0.1667}}
  >{\raggedright\arraybackslash}p{(\linewidth - 10\tabcolsep) * \real{0.1667}}
  >{\raggedright\arraybackslash}p{(\linewidth - 10\tabcolsep) * \real{0.1667}}
  >{\raggedright\arraybackslash}p{(\linewidth - 10\tabcolsep) * \real{0.1667}}@{}}
\caption{Total energies and gallery widths for pristine and C12M
bilayers in AA versus AB stacking.}\tabularnewline
\toprule\noalign{}
\begin{minipage}[b]{\linewidth}\raggedright
system
\end{minipage} & \begin{minipage}[b]{\linewidth}\raggedright
E(AA) / eV
\end{minipage} & \begin{minipage}[b]{\linewidth}\raggedright
E(AB) / eV
\end{minipage} & \begin{minipage}[b]{\linewidth}\raggedright
preference
\end{minipage} & \begin{minipage}[b]{\linewidth}\raggedright
d(AA) / A
\end{minipage} & \begin{minipage}[b]{\linewidth}\raggedright
d(AB) / A
\end{minipage} \\
\midrule\noalign{}
\endfirsthead
\toprule\noalign{}
\begin{minipage}[b]{\linewidth}\raggedright
system
\end{minipage} & \begin{minipage}[b]{\linewidth}\raggedright
E(AA) / eV
\end{minipage} & \begin{minipage}[b]{\linewidth}\raggedright
E(AB) / eV
\end{minipage} & \begin{minipage}[b]{\linewidth}\raggedright
preference
\end{minipage} & \begin{minipage}[b]{\linewidth}\raggedright
d(AA) / A
\end{minipage} & \begin{minipage}[b]{\linewidth}\raggedright
d(AB) / A
\end{minipage} \\
\midrule\noalign{}
\endhead
\bottomrule\noalign{}
\endlastfoot
pristine bilayer & -111.6464 & -111.7023 & AB by 56 meV & 3.768 &
3.417 \\
C12Cr & -118.6144 & -118.0216 & AA by 593 meV & 3.299 & 3.493 \\
C12Mo & -119.6737 & -119.2773 & AA by 396 meV & 3.640 & 3.780 \\
C12W & -118.8285 & -118.3527 & AA by 476 meV & 3.634 & 3.730 \\
\end{longtable}

Pristine bilayer graphene prefers Bernal stacking by 56 meV per sqrt(3)
cell. With any group-6 metal in the gallery the preference reverses, to
between 400 and 600 meV in favour of AA -- an order of magnitude larger
than the intrinsic Bernal preference it has to overcome. In every case
the AB gallery is also wider than the AA one, by 0.10-0.19 A, because
the metal cannot achieve eta-6 coordination to both sheets in AB and is
correspondingly less well bound.

Referenced instead to each stacking's own pristine bilayer, chromium
binds 0.649 eV more strongly in AA than in AB; the difference between
that figure and the 593 meV above is exactly the 56 meV intrinsic Bernal
preference, as it should be.

The conclusion is therefore stronger than ``the stacking penalty is
affordable''. Bis-hexahapto binding actively drives the AA registry,
overcoming the intrinsic Bernal preference by roughly an order of
magnitude. The structural prediction that follows -- AA stacking
wherever the metal is present -- is a robust one rather than a marginal
energetic accident.

\begin{figure}
\centering
\pandocbounded{\includegraphics[keepaspectratio,alt={Figure 2: Band dispersion and gallery midplane density for ionic C12Ca versus covalent C12Cr. (a,b) Fixed-density band paths; linewidth of EF-crossing bands scales with the interstitial/gallery ratio I/G within +/-0.15 eV of EF. (c,d) Atom-masked midplane density for the near-EF gallery state with highest I/G (inset: planar-averaged z profile).}]{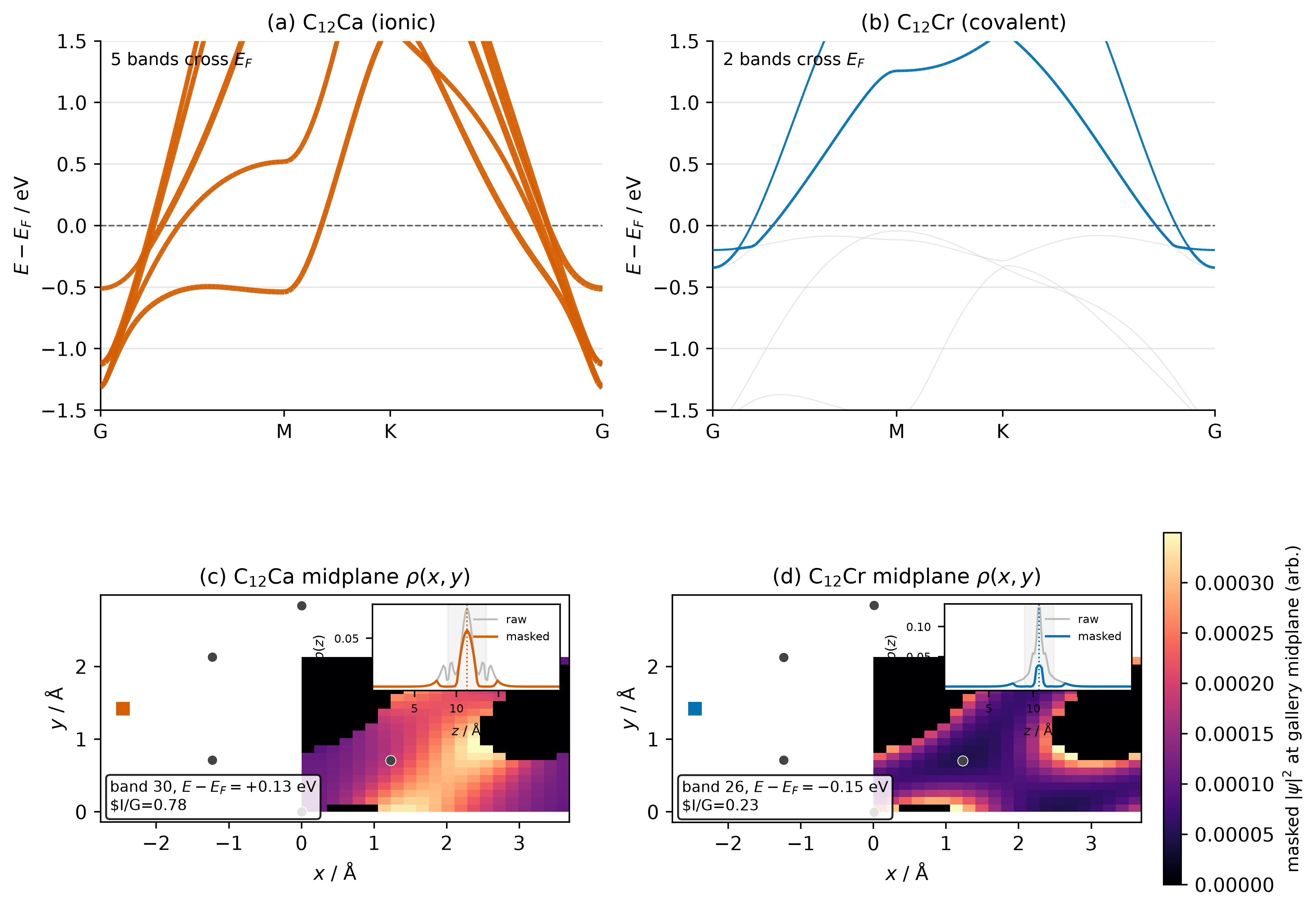}}
\caption*{Figure 2: Band dispersion and gallery midplane density for
ionic C12Ca versus covalent C12Cr. (a,b) Fixed-density band paths;
linewidth of EF-crossing bands scales with the interstitial/gallery
ratio I/G within +/-0.15 eV of EF. (c,d) Atom-masked midplane density
for the near-EF gallery state with highest I/G (inset: planar-averaged z
profile).}
\end{figure}

\begin{figure}
\centering
\pandocbounded{\includegraphics[keepaspectratio,alt={Figure 3: Interstitial fraction of gallery spectral weight at EF (version-2 metric). (a) Same-cell comparison: ionic C12Ca and C12Li versus group-6 C12M. (b) Raw gallery weight versus atom-masked interstitial weight for the same systems.}]{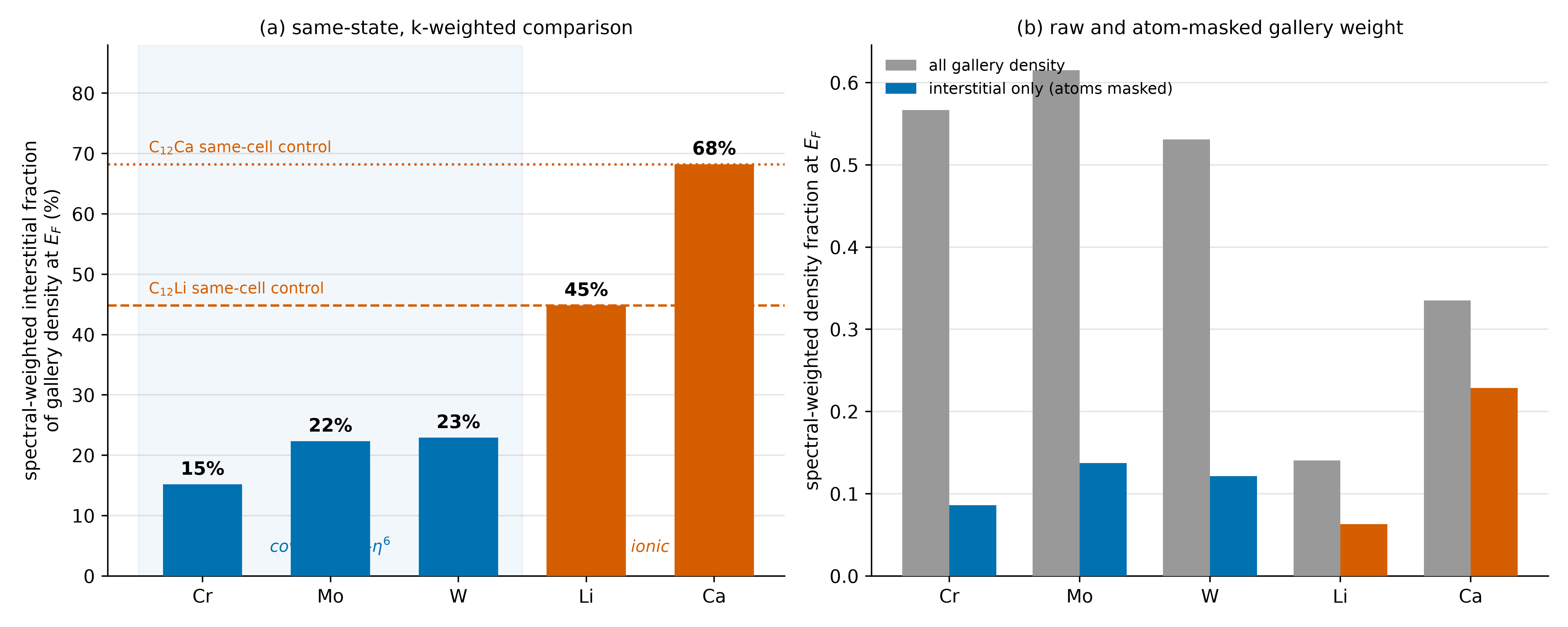}}
\caption*{Figure 3: Interstitial fraction of gallery spectral weight at
EF (version-2 metric). (a) Same-cell comparison: ionic C12Ca and C12Li
versus group-6 C12M. (b) Raw gallery weight versus atom-masked
interstitial weight for the same systems.}
\end{figure}

\begin{figure}
\centering
\pandocbounded{\includegraphics[keepaspectratio,alt={Figure 4: Projected density of states near EF (+/-0.15 eV window), summed over the bound s/p/d projector channels each PAW dataset provides. Ionic controls and bulk CaC6 are carbon dominated; all three group-6 bilayers are metal dominated (about 73 percent), and within that metal weight 89-98 percent is d.~Ca and Li have no bound d projector, so their metal fractions are lower bounds (marked \textgreater).}]{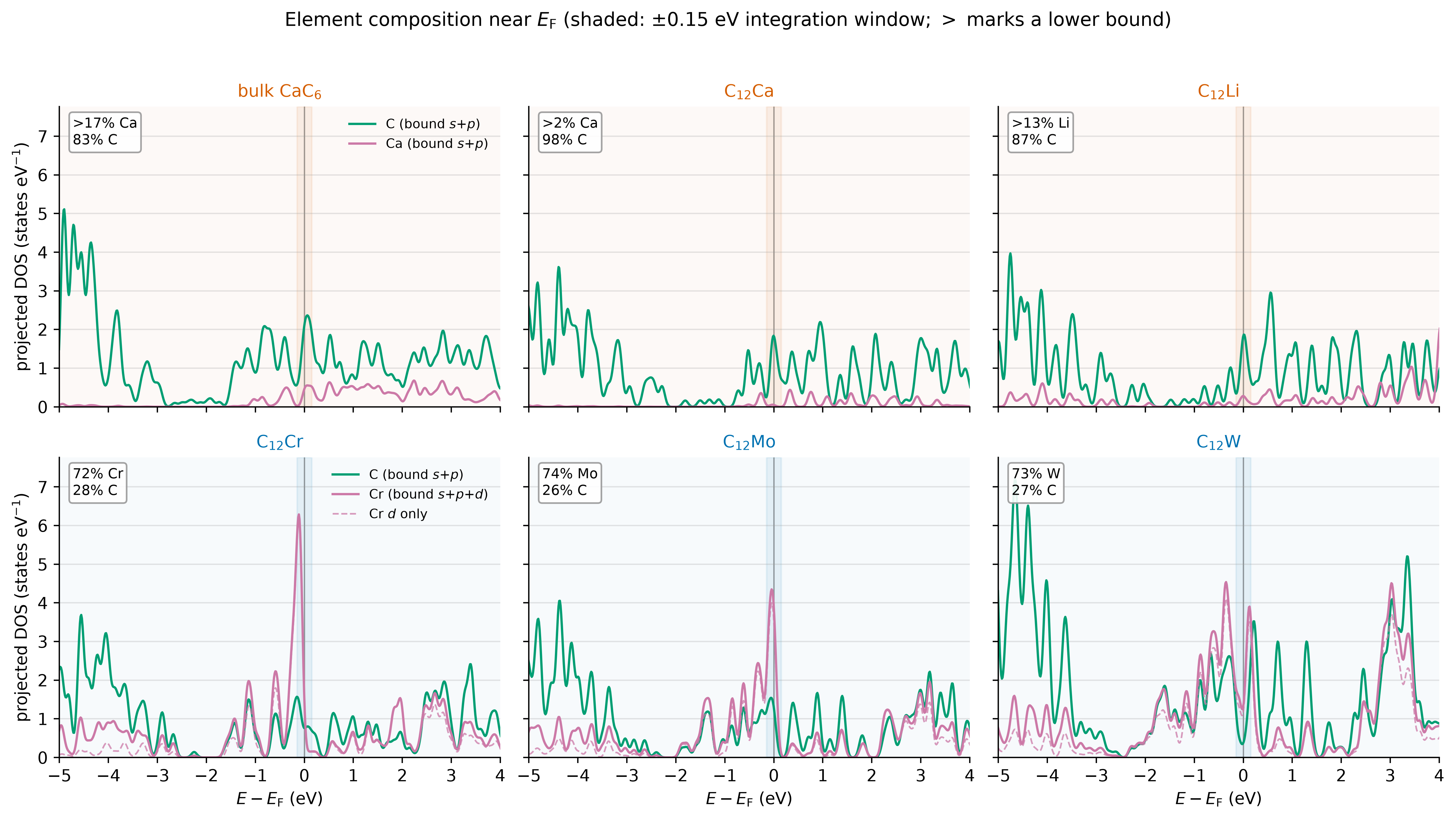}}
\caption*{Figure 4: Projected density of states near EF (+/-0.15 eV
window), summed over the bound s/p/d projector channels each PAW dataset
provides. Ionic controls and bulk CaC6 are carbon dominated; all three
group-6 bilayers are metal dominated (about 73 percent), and within that
metal weight 89-98 percent is d.~Ca and Li have no bound d projector, so
their metal fractions are lower bounds (marked \textgreater).}
\end{figure}

\section{2. Fermi-level character separates ionic controls from
group-6
bilayers}\label{fermi-level-character-separates-ionic-controls-from-group-6-bilayers}

This is the central electronic-structure comparison. All bilayer systems
below share the same sqrt(3) cell at one metal per gallery, are refined
on the production k mesh to a maximum residual force of 0.02 eV/A, and
are analysed with an identical protocol; only the intercalant differs.
For every state within 1.5 eV of the Fermi level we integrate
pseudo-wavefunction density in the central 45\% of the gallery (G\_i)
and the fraction surviving masking of 1.30 A spheres around every
nucleus (I\_i). The primary observable is the k- and occupation-weighted
ratio R = sum(q\_i I\_i) / sum(q\_i G\_i), with q\_i the irreducible k
weight times a Gaussian kernel of sigma = 0.10 eV centred on E\_F. The
absolute weighted interstitial integral A = sum(q\_i I\_i) is reported
separately. Bulk CaC6 is analysed under the same protocol as a
construct-positive benchmark; it is not substituted for the same-cell
bilayer controls in the headline comparison, and absolute values of A
are not compared across different unit cells.

\begin{longtable}[]{@{}
  >{\raggedright\arraybackslash}p{(\linewidth - 8\tabcolsep) * \real{0.2000}}
  >{\raggedright\arraybackslash}p{(\linewidth - 8\tabcolsep) * \real{0.2000}}
  >{\raggedright\arraybackslash}p{(\linewidth - 8\tabcolsep) * \real{0.2000}}
  >{\raggedright\arraybackslash}p{(\linewidth - 8\tabcolsep) * \real{0.2000}}
  >{\raggedright\arraybackslash}p{(\linewidth - 8\tabcolsep) * \real{0.2000}}@{}}
\caption{Interlayer-character metrics near the Fermi level for ionic
controls and group-6 C12M bilayers.}\tabularnewline
\toprule\noalign{}
\begin{minipage}[b]{\linewidth}\raggedright
system
\end{minipage} & \begin{minipage}[b]{\linewidth}\raggedright
bonding
\end{minipage} & \begin{minipage}[b]{\linewidth}\raggedright
R (interstitial / gallery)
\end{minipage} & \begin{minipage}[b]{\linewidth}\raggedright
A
\end{minipage} & \begin{minipage}[b]{\linewidth}\raggedright
raw gallery weight G
\end{minipage} \\
\midrule\noalign{}
\endfirsthead
\toprule\noalign{}
\begin{minipage}[b]{\linewidth}\raggedright
system
\end{minipage} & \begin{minipage}[b]{\linewidth}\raggedright
bonding
\end{minipage} & \begin{minipage}[b]{\linewidth}\raggedright
R (interstitial / gallery)
\end{minipage} & \begin{minipage}[b]{\linewidth}\raggedright
A
\end{minipage} & \begin{minipage}[b]{\linewidth}\raggedright
raw gallery weight G
\end{minipage} \\
\midrule\noalign{}
\endhead
\bottomrule\noalign{}
\endlastfoot
bulk CaC6 & ionic (benchmark) & 0.70 & 0.089 & 0.128 \\
C12Ca & ionic (same-cell control) & 0.68 & 0.229 & 0.335 \\
C12Li & ionic (same-cell control) & 0.45 & 0.063 & 0.140 \\
C12Cr & covalent & 0.15 & 0.086 & 0.567 \\
C12Mo & covalent & 0.22 & 0.137 & 0.615 \\
C12W & covalent & 0.23 & 0.122 & 0.531 \\
\end{longtable}

\subsection{Composition, not
depletion}\label{composition-not-depletion}

The ionic same-cell controls retain a larger interstitial
\emph{fraction} R than any group-6 bilayer: C12Ca (0.68) and C12Li
(0.45) against 0.15-0.23 for Cr, Mo and W. That ordering is not driven
by a deficit of gallery density in the covalent systems. Raw gallery
weight at E\_F is 0.53-0.62 for the group-6 metals against 0.13-0.14 for
the ionic controls, and the absolute interstitial weight A is likewise
higher for every group-6 system than for C12Li (0.063). The contrast is
therefore one of composition: a smaller fraction of a much larger
gallery spectral weight survives atom masking because the density is
metal-d-derived and atom-centred rather than carbon-p-derived and
interstitial-like.

Mesh convergence (12x12 to 18x18), mask-radius and gallery-width
sensitivities are documented in the Supporting Information. At the
frozen baseline parameters every group-6 R lies at least 0.05 below
R(C12Li). A prespecified sensitivity envelope test comparing each
system's worst case against the control's best case does not pass
because the envelopes overlap when the atom-mask radius is varied; a
post-hoc check at matched parameter settings finds R(group-6)
\textless{} R(C12Li) at every common setting. That matched-parameter
ordering is reported in the Supporting Information and labelled as such;
parameter envelopes are shown as envelopes, not as error bars.

\subsection{Orbital projection at
E\_F}\label{orbital-projection-at-e_f}

Projected density of states provides a decomposition that does not
reduce to the masking operation used to define R. We integrate element-
and l-resolved PDOS in \textbar E - E\_F\textbar{} \textless= 0.15 eV
and normalise by element, summing for each element the bound s, p and d
projector channels its PAW dataset provides.

The normalisation is by element rather than by orbital channel for a
reason that is easy to get wrong. GPAW's projected density of states is
built from \emph{bound} PAW projector functions only. In the default
datasets calcium's d projector is unbound and lithium has no d projector
at all, so a Ca-d or Li-d PDOS is identically zero over the entire
energy range by construction, whereas Cr, Mo and W have bound d and
report real weight. A metal-d against carbon-p comparison would
therefore set a channel that cannot be nonzero against one that can.
Normalising by element removes that asymmetry. What survives is a stated
limitation rather than a hidden one: a bound-projector scheme still
cannot see Ca-d hybridisation, so the metal fraction quoted for the
ionic controls is a lower bound, marked below with a greater-than sign.

\begin{longtable}[]{@{}lllll@{}}
\caption{Element composition of states within +/- 0.15 eV of EF from
projected density of states. Values for Ca and Li are lower bounds (no
bound d projector).}\tabularnewline
\toprule\noalign{}
system & metal (\%) & carbon (\%) & d as \% of metal & classification \\
\midrule\noalign{}
\endfirsthead
\toprule\noalign{}
system & metal (\%) & carbon (\%) & d as \% of metal & classification \\
\midrule\noalign{}
\endhead
\bottomrule\noalign{}
\endlastfoot
bulk CaC6 & \textgreater{} 17 & 83 & n/a & carbon dominated \\
C12Ca & \textgreater{} 2 & 98 & n/a & carbon dominated \\
C12Li & \textgreater{} 13 & 87 & n/a & carbon dominated \\
C12Cr & 72 & 28 & 98 & metal dominated \\
C12Mo & 74 & 26 & 91 & metal dominated \\
C12W & 73 & 27 & 89 & metal dominated \\
\end{longtable}

The ionic references and bulk CaC6 are carbon dominated at the Fermi
level; all three group-6 bilayers are metal dominated at comparable
percentages, and within that metal weight 89-98 per cent is d.~The
separation is categorical even at its narrowest: a lower bound of 17 per
cent for the benchmark against 72 per cent for chromium. Fine trends
among Cr, Mo and W within the group-6 series are not resolved at this
level. This is evidence about orbital character, not by itself a proof
that electron-phonon coupling or a superconducting transition is absent.

\subsection{The benchmark's interlayer band, identified without the
mask}\label{the-benchmarks-interlayer-band-identified-without-the-mask}

Both measures above are built from atomic spheres -- the interstitial
fraction R masks them, the projected density of states projects onto
them -- so neither can validate the other independently. A third test
uses the eigenvalues alone. A nearly free-electron interlayer state is
delocalised in the gallery and must therefore disperse in three
dimensions: a parabolic minimum at the zone centre with a light and
nearly \emph{isotropic} effective mass. A carbon pi band in a layered
crystal cannot do this, being light in plane and heavy across the
layers.

In bulk CaC6 exactly three bands cross E\_F with their minimum at Gamma.
Two are strongly anisotropic (m* = 0.20 in plane against 0.79 across the
layers, a ratio of 3.9). The third, 1.27 eV below E\_F at Gamma, has m*
= 0.49 in plane and 0.47 across the layers -- a ratio of 1.03, and a
mass of about half the free-electron value. That is the interlayer band,
partially occupied, and it is identified here with no reference to where
any density sits -- which is also the limit of what this test shows.
Nearly free-electron, light and isotropic is the dispersion half of the
criterion of Csányi and co-workers {[}3{]}; the spatial half, that the
state is well separated from the carbon sheets, is supplied by the
atom-masked measure above (R = 0.70 for bulk CaC6), not by the fit.
Taken together the two discharge the construct-validity condition of the
frozen specification. The mask-based per-state rule names a different
band from the dispersion rule -- the two optimise different quantities,
and both bands carry high interstitial character near E\_F; the
cross-check is in the Supporting Information.

All six bilayer systems and bulk CaC6 are metallic on dense uniform
meshes with a partial-occupation test. The operative question is
therefore not whether a Fermi surface exists but what carries it: for
the group-6 bilayers, metal d states; for the ionic controls and bulk
CaC6, carbon states with a substantially larger interstitial fraction R.

\subsection{Band dispersion near
E\_F}\label{band-dispersion-near-e_f}

Fixed-density band structures on the production geometries confirm that
both same-cell controls and all group-6 bilayers are metallic. Figure 2
compares C12Ca with C12Cr on identical high-symmetry paths. Panel (a,b):
both exhibit bands crossing E\_F; linewidth of the highlighted crossings
scales with the interstitial/gallery ratio I/G sampled along the path
(within \$\pm\$0.15 eV of E\_F). C12Ca shows several dispersive
crossings whose orbital composition is carbon p dominated (above) and
carries a much larger path-averaged I/G; C12Cr crosses E\_F on fewer
segments with thinner linewidth, consistent with atom-centred metal d
states rather than a carbon-derived interlayer band. Panel (c,d):
atom-masked \(|\psi|^2\) at the gallery midplane for the near-E\_F
gallery state with highest I/G in each system (inset: planar-averaged z
profile). C12Ca peaks between the sheets away from the nuclei; C12Cr
weight concentrates on the intercalant. Figures 3 and 4 report the
complementary interstitial-fraction and PDOS comparisons for the full
same-cell series. A one-dimensional path is not a complete map of the
Fermi surface; the band plot and midplane maps are included to show that
metallicity is not in dispute and that the ionic-covalent contrast rests
on orbital assignment, interstitial character, and spatial topology, not
on the presence of states at E\_F.

\section{3. C12Cr is a non-magnetic METAL, not a closed-shell
semiconductor}\label{c12cr-is-a-non-magnetic-metal-not-a-closed-shell-semiconductor}

At one metal per sqrt(3) gallery -- the coverage matching bulk CaC6's
gallery density -- the relaxed gallery width is 3.299 A, essentially
unchanged from an estimate built from the 2.14 A Cr-C distance in
bis(benzene)chromium (3.20 A). This is \emph{narrower} than graphite's
3.35 A: the bis-hexahapto bond draws the sheets together.

A ferromagnetic calculation seeded at 4 muB relaxes to zero moment and
to the non-magnetic energy to four decimal places (-118.6144 eV in
both). No magnetic solution exists for this system to find.

\emph{The system is nevertheless metallic.} Determined by band index
with a partial-occupation test, on a dense uniform mesh containing the K
point, C12Cr has no gap: bands cross the Fermi level. The same is true
of C12Mo and C12W.

This distinction matters, and it revises the picture inherited from the
literature. A previous study {[}13{]} reports a 0.24 eV gap for C12Cr,
attributed to d-pi interaction driving each C6CrC6 unit to an
18-electron closed shell. Our calculations reproduce the \emph{bonding}
consequence of that picture -- there is no magnetic solution, as a
closed shell implies -- but not the \emph{gap}. Absence of a moment does
not require a gap: a non-magnetic metal with modest density of states at
E\_F behaves exactly this way.

The discrepancy is traceable to Brillouin-zone sampling rather than to
chemistry. Measured on the self-consistent integration mesh, an apparent
gap of 0.27 eV on a 6x6 mesh falls to 0.014 eV on 18x18 and to zero when
determined by band index rather than by position relative to E\_F
(Supporting Information). Because K sits at (1/3, 1/3), meshes whose
divisions are not multiples of three do not sample it at all, and a band
edge there is invisible. The 18-electron \emph{electron count} is not in
question; its identification with a semiconducting gap is.

\begin{figure}
\centering
\pandocbounded{\includegraphics[keepaspectratio,alt={Figure 5: Constrained-moment energy E(M) relative to M = 0. Tungsten (C12W) shows a weak itinerant minimum near 0.5 muB; chromium (C6Cr) has no stable high-moment solution under this Hamiltonian.}]{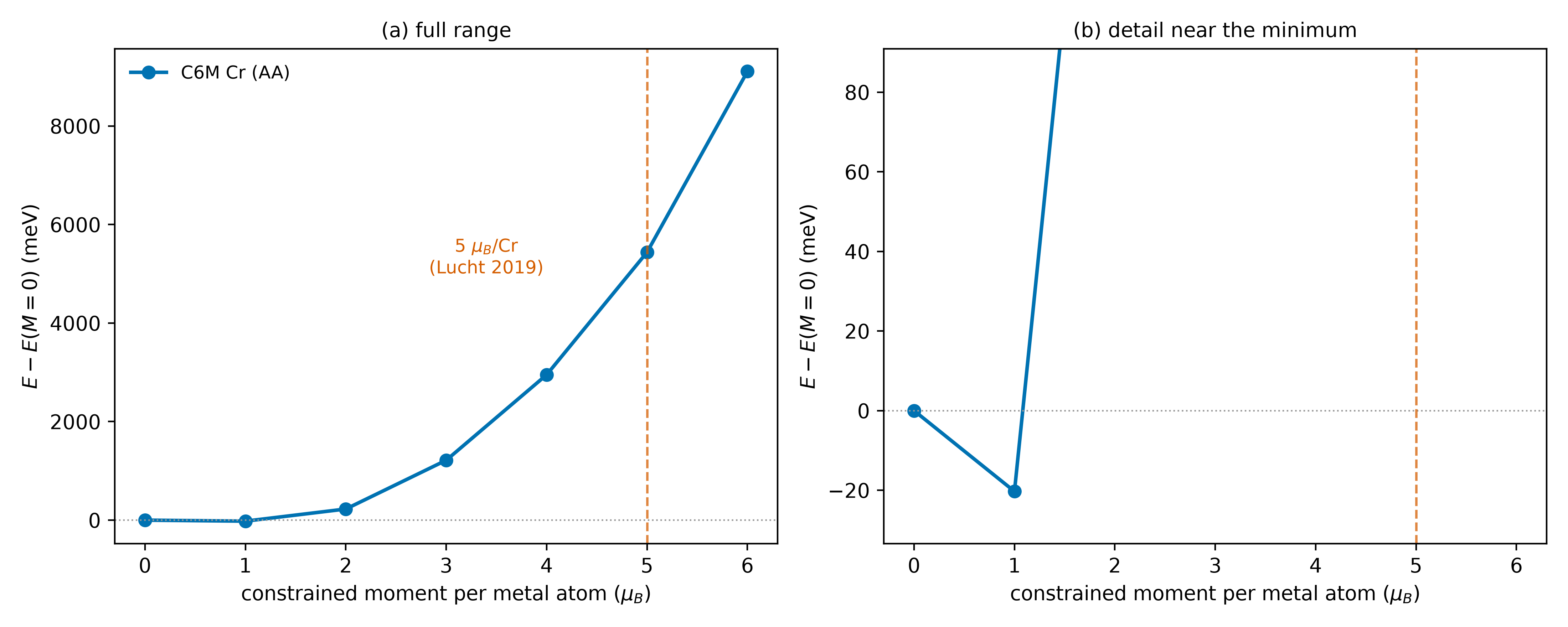}}
\caption*{Figure 5: Constrained-moment energy E(M) relative to M = 0.
Tungsten (C12W) shows a weak itinerant minimum near 0.5 muB; chromium
(C6Cr) has no stable high-moment solution under this Hamiltonian.}
\end{figure}

\section{4. C6Cr: a magnetic instability pre-empts a Peierls
instability}\label{c6cr-a-magnetic-instability-pre-empts-a-peierls-instability}

At two metals per sqrt(3) gallery the metal atoms form a honeycomb
sublattice with a nearest-neighbour separation of 2.454 A -- shorter
than the 2.498 A nearest-neighbour distance in bcc chromium metal. The
gallery therefore contains a metal-metal bonded chromium layer, not an
array of isolated sandwich units.

Three magnetic configurations were relaxed:

\begin{longtable}[]{@{}lllll@{}}
\caption{Magnetic and structural comparison of ferromagnetic,
antiferromagnetic and non-magnetic C6Cr.}\tabularnewline
\toprule\noalign{}
state & E / eV & dE vs FM & moment & Cr-Cr shells / A \\
\midrule\noalign{}
\endfirsthead
\toprule\noalign{}
state & E / eV & dE vs FM & moment & Cr-Cr shells / A \\
\midrule\noalign{}
\endhead
\bottomrule\noalign{}
\endlastfoot
FM & -125.8170 & 0 & 1.79 muB (\textasciitilde0.9/Cr) & 2.454, 2.454,
2.472 \\
AFM & -125.7360 & +81 meV & \textasciitilde0 & 2.350, 2.517, 2.517 \\
NM & -125.7357 & +81 meV & 0 & 2.351, 2.516, 2.516 \\
\end{longtable}

The antiferromagnetic and non-magnetic calculations converge to the same
solution, agreeing to 0.3 meV in energy and 0.001 A in every metal-metal
distance. The bond alternation -- one short contact at 2.35 A and two
long at 2.52 A -- is therefore a property of the non-magnetic state: a
Peierls-like distortion that opens a gap at the Fermi level. The
ferromagnetic state instead develops a moment and retains a uniform
honeycomb, and wins by 81 meV.

Constrained-moment calculations decompose that 81 meV: at the fixed
distorted geometry, magnetism is worth about 20 meV, with the remaining
\textasciitilde61 meV coming from the structure relaxing back to a
uniform honeycomb once the moment is present. The moment forms first,
and the distortion then becomes unfavourable.

\subsection{The distortion is not an artefact of the unit
cell}\label{the-distortion-is-not-an-artefact-of-the-unit-cell}

A sqrt(3)xsqrt(3) cell contains only two metal atoms, so exactly one
bond-alternation pattern is available to it and observing one there is
weak evidence. We therefore repeated the calculation in a doubled cell
containing four chromium atoms, starting from a uniform honeycomb at the
relaxed gallery width with a small random in-plane displacement (0.05 A
rms) applied to the metals only -- enough to break the symmetry without
imposing a pattern.

The distortion appears spontaneously, and is larger than in the small
cell:

\begin{longtable}[]{@{}
  >{\raggedright\arraybackslash}p{(\linewidth - 6\tabcolsep) * \real{0.2500}}
  >{\raggedright\arraybackslash}p{(\linewidth - 6\tabcolsep) * \real{0.2500}}
  >{\raggedright\arraybackslash}p{(\linewidth - 6\tabcolsep) * \real{0.2500}}
  >{\raggedright\arraybackslash}p{(\linewidth - 6\tabcolsep) * \real{0.2500}}@{}}
\caption{Cr--Cr bond alternation in the non-magnetic state for two cell
choices.}\tabularnewline
\toprule\noalign{}
\begin{minipage}[b]{\linewidth}\raggedright
cell
\end{minipage} & \begin{minipage}[b]{\linewidth}\raggedright
metals
\end{minipage} & \begin{minipage}[b]{\linewidth}\raggedright
max Cr-Cr spread
\end{minipage} & \begin{minipage}[b]{\linewidth}\raggedright
pattern
\end{minipage} \\
\midrule\noalign{}
\endfirsthead
\toprule\noalign{}
\begin{minipage}[b]{\linewidth}\raggedright
cell
\end{minipage} & \begin{minipage}[b]{\linewidth}\raggedright
metals
\end{minipage} & \begin{minipage}[b]{\linewidth}\raggedright
max Cr-Cr spread
\end{minipage} & \begin{minipage}[b]{\linewidth}\raggedright
pattern
\end{minipage} \\
\midrule\noalign{}
\endhead
\bottomrule\noalign{}
\endlastfoot
sqrt(3)xsqrt(3) & 2 & 0.165 A & single pattern available \\
2x1 supercell & 4 & 0.248 A & two inequivalent environments (0.248,
0.248, 0.092, 0.100 A) \\
\end{longtable}

Given the freedom to choose, the system adopts a distortion that the
smaller cell cannot represent: the four metals fall into two
inequivalent environments rather than pairing uniformly. A cell-imposed
artefact would have disappeared when the constraint was lifted; instead
the amplitude grew by half and acquired additional structure. The bond
alternation in the non-magnetic state is a real instability, and the
sqrt(3) cell understates it.

\subsection{Comparison with previous
work}\label{comparison-with-previous-work}

A previous study {[}12{]} of first-row transition metals in this
geometry reports an antiferromagnetic ground state with a moment of
about 5 muB per chromium. The present calculations disagree on both
counts, and the constrained-moment curve shows why the disagreement is
not a matter of near-degenerate states being ordered differently:

\begin{longtable}[]{@{}lll@{}}
\caption{Constrained-moment energies for C6Cr relative to the
non-magnetic state.}\tabularnewline
\toprule\noalign{}
constrained moment & per Cr & E - E(M=0) \\
\midrule\noalign{}
\endfirsthead
\toprule\noalign{}
constrained moment & per Cr & E - E(M=0) \\
\midrule\noalign{}
\endhead
\bottomrule\noalign{}
\endlastfoot
0 & 0 & 0 \\
2 muB & 1.0 & -20 meV (minimum) \\
4 muB & 2.0 & +226 meV \\
6 muB & 3.0 & +1.22 eV \\
8 muB & 4.0 & +2.95 eV \\
10 muB & 5.0 & +5.44 eV \\
\end{longtable}

The energy rises superlinearly beyond a shallow minimum at 1 muB per
chromium. A configuration with 5 muB per chromium -- the value reported
previously -- lies 5.44 eV above the non-magnetic state and 5.46 eV
above the ferromagnetic ground state. It is not a solution of this
Hamiltonian, and no choice of dispersion correction, Hubbard U, or
smearing width closes a deficit of that size.

Two independent considerations point the same way. First, a moment of
\textasciitilde5 muB approaches the free-atom value (6 muB, reproduced
exactly here for an isolated Cr atom) and is not physically available at
a metal-metal separation shorter than in chromium metal, where the
moment is only \textasciitilde0.6 muB. Second, the earlier work used a
localised numerical basis; in our tested double-zeta-polarised LCAO
basis at these gallery spacings the overlap matrix loses
positive-definiteness, which is why the present work uses plane waves. A
localised-basis calculation that remains stable under these conditions
may be using a basis too small to describe metal-pi hybridisation fully,
and an under-hybridised d manifold retains an inflated, atom-like
moment.

\begin{figure}
\centering
\pandocbounded{\includegraphics[keepaspectratio,alt={Figure 6: Formation energies of ordered C12M phases. (a) Relative to bulk metal (positive: metastable against segregation) and to isolated atoms (negative: bound under metal-vapour conditions). (b) Metal cohesive energies implied by the gap between the two references, compared with experiment.}]{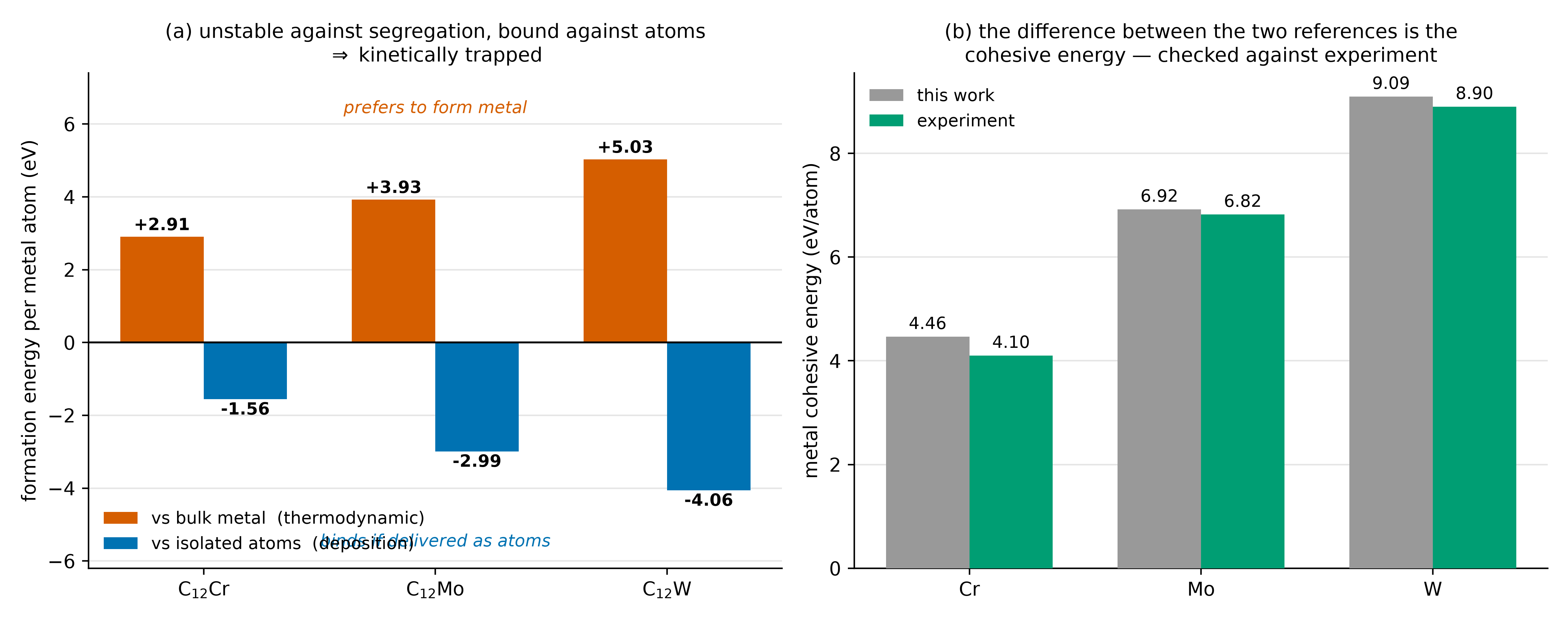}}
\caption*{Figure 6: Formation energies of ordered C12M phases. (a)
Relative to bulk metal (positive: metastable against segregation) and to
isolated atoms (negative: bound under metal-vapour conditions). (b)
Metal cohesive energies implied by the gap between the two references,
compared with experiment.}
\end{figure}

\section{5. Ordered group-6 intercalation is kinetically trapped, not
thermodynamically
stable}\label{ordered-group-6-intercalation-is-kinetically-trapped-not-thermodynamically-stable}

Formation energies per intercalated metal atom, referenced two
independent ways:

\begin{longtable}[]{@{}llll@{}}
\caption{Formation energies of C12M and C6Cr relative to bulk metal and
isolated atoms.}\tabularnewline
\toprule\noalign{}
system & vs bulk metal & vs isolated atom & gallery / A \\
\midrule\noalign{}
\endfirsthead
\toprule\noalign{}
system & vs bulk metal & vs isolated atom & gallery / A \\
\midrule\noalign{}
\endhead
\bottomrule\noalign{}
\endlastfoot
C12Cr AA & +2.906 eV & -1.558 eV & 3.299 \\
C12Mo AA & +3.927 eV & -2.992 eV & 3.640 \\
C12W AA & +5.032 eV & -4.063 eV & 3.634 \\
C6Cr AA & +2.789 eV & -1.675 eV & 3.599 \\
\end{longtable}

The two references answer different questions and, taken together,
resolve an apparent paradox in the experimental literature.

Against bulk metal, every ordered phase is unstable by 2.8-5.0 eV per
metal atom. An ordered group-6 sublattice in a graphene gallery is
therefore not a thermodynamic minimum: the metal prefers to segregate
and form metal. This bears directly on the proposal that these materials
might be prepared with the metal atoms ``arranged in a specific order''
-- the ordered arrangement is not what thermodynamics selects.

Against isolated atoms, binding is favourable by 1.6-4.1 eV. This is the
reference appropriate to metal-vapour synthesis, where atoms rather than
bulk metal are delivered to the surface, and it explains why the
experiment succeeds: the intercalated phase is kinetically trapped. Once
formed it is bound relative to the atoms that made it, while remaining
metastable with respect to clustering. That is consistent with the
experimental observation that these metals are mobile on graphitic
surfaces and are known cluster-formers.

The two references give opposite orderings, and the difference between
them is exactly the metal cohesive energy, which rises steeply down
group 6. Extracting it provides an independent check of the reference
states:

\begin{longtable}[]{@{}lll@{}}
\caption{Computed metal cohesive energies compared with
experiment.}\tabularnewline
\toprule\noalign{}
metal & cohesive energy, this work & experiment \\
\midrule\noalign{}
\endfirsthead
\toprule\noalign{}
metal & cohesive energy, this work & experiment \\
\midrule\noalign{}
\endhead
\bottomrule\noalign{}
\endlastfoot
Cr & 4.46 eV & 4.10 eV \\
Mo & 6.92 eV & 6.82 eV \\
W & 9.10 eV & 8.90 eV \\
\end{longtable}

The agreement (slightly overbound, as expected of PBE) confirms that
both the bulk and atomic references are sound.

\subsection{The bulk-referenced trend reproduces the measured
reactivity
ordering}\label{the-bulk-referenced-trend-reproduces-the-measured-reactivity-ordering}

Ordered by instability against segregation -- smallest positive
formation energy first -- the series runs Cr (+2.91) \textless{} Mo
(+3.93) \textless{} W (+5.03) eV: chromium is the least unstable and
therefore the most likely to form an intercalated phase. That ordering
of thermodynamic metastabilities corresponds to the experimental
reactivity ordering Cr \textgreater{} Mo \textgreater{} W for the
group-6 hexacarbonyl reagents (chromium most reactive, tungsten least)
measured on graphene and PPG graphene-nanoplatelet thin films, reported
in {[}9{]} and set out in more detail in the present author's
dissertation {[}10{]}. The agreement is notable because the experimental
ordering could in principle have been governed entirely by reagent
photochemistry -- the relative ease of photolysing Cr(CO)6, Mo(CO)6 and
W(CO)6 -- rather than by metal-graphene affinity. That the thermodynamic
ordering matches suggests the observed reactivity trend has a
thermodynamic component and is not purely a property of the precursors.

These formation energies describe an ideal ordered sublattice at 0 K and
do not, by themselves, set a room-temperature hop rate. Chromium is
observed to move between macroscopic graphene sheets at 40-195 \(\mu\)m
s-1 at room temperature in metal-vapour-synthesis devices ({[}10{]}, the
present author's dissertation), and reconciling that mobility with the
metastability reported here requires migration-barrier calculations that
we have not performed; the two observations are not in contradiction if
long-range order is a local or kinetic product of synthesis rather than
a global thermodynamic minimum.

\emph{Caveat on the tungsten atomic reference:} PBE places the free W
atom in a 5d5 6s1 configuration (6 muB), whereas the experimental ground
state is 5d4 6s2 (4 muB). This is a known deficiency of semi-local
functionals for 5d s-d promotion energies. It shifts the atom-referenced
W value by of order the promotion energy; the bulk-referenced values, on
which the reactivity comparison rests, are unaffected. All three metals
are referenced to the same configuration type, so the trend is more
robust than any individual absolute value.

\subsection{The ordered phases carry no zone-centre lattice
instability}\label{the-ordered-phases-carry-no-zone-centre-lattice-instability}

Metastability against segregation is a statement about energy, not about
whether the ordered lattice will hold its shape. We therefore computed
zone-centre phonons for the three C12M bilayers by finite displacement
(delta = 0.01 A, both signs, all 3N degrees of freedom) on the
production geometries, with the same calculator and with D3 contributing
to the force constants. The acceptance criterion was fixed in the
analysis specification before any frequency was computed, because the
quantity that decides the question -- how far the numerical noise floor
sits below the softest real mode -- is exactly the kind of threshold
that invites adjustment after the fact.

None of the three has an imaginary optical mode at the zone centre. The
noise floor, measured as the residual splitting of the three acoustic
branches with the acoustic sum rule switched off, lies well below the
softest optical branch in every case: by a factor of 16 for C12Cr (10
against 163 cm-1), 8 for C12W (18 against 152 cm-1) and 2.6 for C12Mo
(43 against 112 cm-1). An independent check comes free with the
calculation: the highest branch of each system falls at 1583-1606 cm-1,
within 1.7 per cent of graphene's measured in-plane C-C stretch near
1580 cm-1, and nothing in the procedure was fitted to it.

The pristine bilayer, computed in both stackings as a control under
identical settings, separates cleanly and in the direction Section 1
predicts. Bernal AB, the experimental ground state, has no imaginary
mode: its two softest optical branches, at 22 and 24 cm-1, are the
interlayer shear doublet, in the range of the \textasciitilde31 cm-1
shear mode measured in bilayer graphene. In AA the same doublet splits
and one member goes imaginary, at -29 cm-1, 99.99 per cent in-plane with
the two sheets moving in antiphase (direction cosine -1.000). That is
the sliding coordinate which carries AA into AB, and its sign flips
between the two stackings exactly as the 56 meV Bernal preference
requires.

The comparison also disposes of the obvious worry about a soft-mode
result. The noise floor is essentially the same in the two stackings --
16 cm-1 in AB against 19 cm-1 in AA -- so it is a property of the
settings, not of the structure, and it cannot be what makes AA unstable
and AB stable. Individually these low frequencies are not well resolved:
AB's softest branch clears its own noise floor by only a factor of 1.4,
so the shear frequency itself carries a large relative uncertainty. What
is robust is the difference. A swing of about 50 cm-1 between -29 and
+22 cm-1 is far larger than either noise floor, so the change of sign
survives even though neither magnitude is precise.

The control therefore does double duty. It confirms that the method
finds a genuine soft mode where one is known to exist, which is what
gives weight to its absence in the intercalated systems; and it shows
directly that the metal is what holds the AA registry, since the shear
coordinate that is unstable in the empty AA gallery is stable once a
metal occupies it.

A second imaginary branch appears in AA at -159 cm-1, 88 per cent
out-of-plane, where AB has a real branch at +204 cm-1. It is therefore
specific to the stacking rather than an artefact of the settings, and
the numerical explanations are in any case excluded: the in-plane
lattice constant matches graphene's ideal sqrt(3) value to four decimal
places, and central differences are insensitive to the reference
geometry's residual forces at first order. Its out-of-plane character is
what one would expect of a distortion relieving the Pauli repulsion
between eclipsed pi clouds that Section 1 identifies as the origin of
the AA penalty, but we have not established that assignment and do not
rest anything on it.

\emph{What this does and does not establish.} A (1,1,1) supercell
resolves q = 0 of the sqrt(3)xsqrt(3)R30 cell, which is more than the
zone centre of graphene: that cell folds the primitive-cell K point onto
Gamma, so the spectra above already contain the carbon K-point modes.
What they do not contain is the primitive-cell M point, or anything away
from the zone centre of the metal superlattice. These calculations can
therefore exclude a zone-centre instability; they cannot establish
dynamical stability, and we do not claim it. The distinction is not
academic here: the bond alternation reported for non-magnetic C6Cr in
Section 4 is a zone-boundary instability of the metal sublattice in a
different cell again, and a calculation of this kind is blind to it by
construction. A full q mesh requires a 2x2 supercell -- 52 atoms and 312
displacements -- which was beyond the resources available. Full
frequency lists, the noise-floor margins, and the displacement patterns
of every imaginary mode are in the Supporting Information.

\section{6. The closed-shell description holds for Cr and Mo but
fails for
W}\label{the-closed-shell-description-holds-for-cr-and-mo-but-fails-for-w}

At one metal per gallery, chromium and molybdenum behave identically: a
ferromagnetic calculation seeded at 4 muB relaxes to zero moment and to
the non-magnetic energy, to four decimal places in both cases. No
magnetic solution exists. Since molybdenum has not previously been
treated, this establishes that the behaviour is not specific to
chromium.

\emph{Tungsten does not follow.} Its ferromagnetic state carries 0.65
muB and lies 49 meV below non-magnetic. A constrained-moment scan
confirms this is a genuine minimum rather than an artefact of
self-consistency:

\begin{longtable}[]{@{}ll@{}}
\caption{Constrained-moment scan for C12W.}\tabularnewline
\toprule\noalign{}
constrained M / muB & E - E(M=0) / meV \\
\midrule\noalign{}
\endfirsthead
\toprule\noalign{}
constrained M / muB & E - E(M=0) / meV \\
\midrule\noalign{}
\endhead
\bottomrule\noalign{}
\endlastfoot
0 & 0 \\
0.5 & -34.9 (minimum) \\
1.0 & -10.4 \\
1.5 & +128 \\
2.0 & +349 \\
\end{longtable}

The minimum at 0.5 muB brackets the 0.65 muB found by free relaxation,
and the 34.9 meV at fixed geometry plus \textasciitilde14 meV from the
gallery relaxing (3.634 to 3.643 A) accounts for the 49 meV of the full
calculation. The shape -- a shallow minimum with a steep rise beyond
\textasciitilde1 muB -- is that of a weak itinerant magnet.

The moment appears only in AA stacking. In AB, where eta-6 coordination
to both sheets is geometrically unavailable, ferromagnetic and
non-magnetic calculations give identical energies and no moment.

The mechanism is not established, and we decline to assert one. An
obvious candidate -- that the closed-shell gap closes progressively down
the group, allowing a Stoner instability in W that a gapped system could
not support -- is excluded by our own results: the dense-mesh,
band-index analysis finds all three metals metallic, with no gap for Cr,
Mo or W. Metallicity therefore cannot be what distinguishes tungsten. A
Stoner mechanism would instead require an unusually large density of
states at E\_F for W, which is counterintuitive given that 5d exchange
integrals are smaller than 3d. That the moment tracks the availability
of eta-6 coordination (present in AA, absent in AB) suggests a narrow
d-derived band produced by the bonding itself; projected density of
states at E\_F (Section 2) confirms that the group-6 bilayers are
metal-d dominated, which is consistent with this picture though not
sufficient to identify the mechanism uniquely.

\begin{figure}
\centering
\pandocbounded{\includegraphics[keepaspectratio,alt={Figure 7: Coverage dependence across Cr, Mo and W. (a) Gallery width versus coverage; crowding prises Mo/W sheets apart at C6M. (b) Magnetic member of the series inverts with coverage: Cr is magnetic at C6M but not C12M; W is magnetic at C12M but not C6M.}]{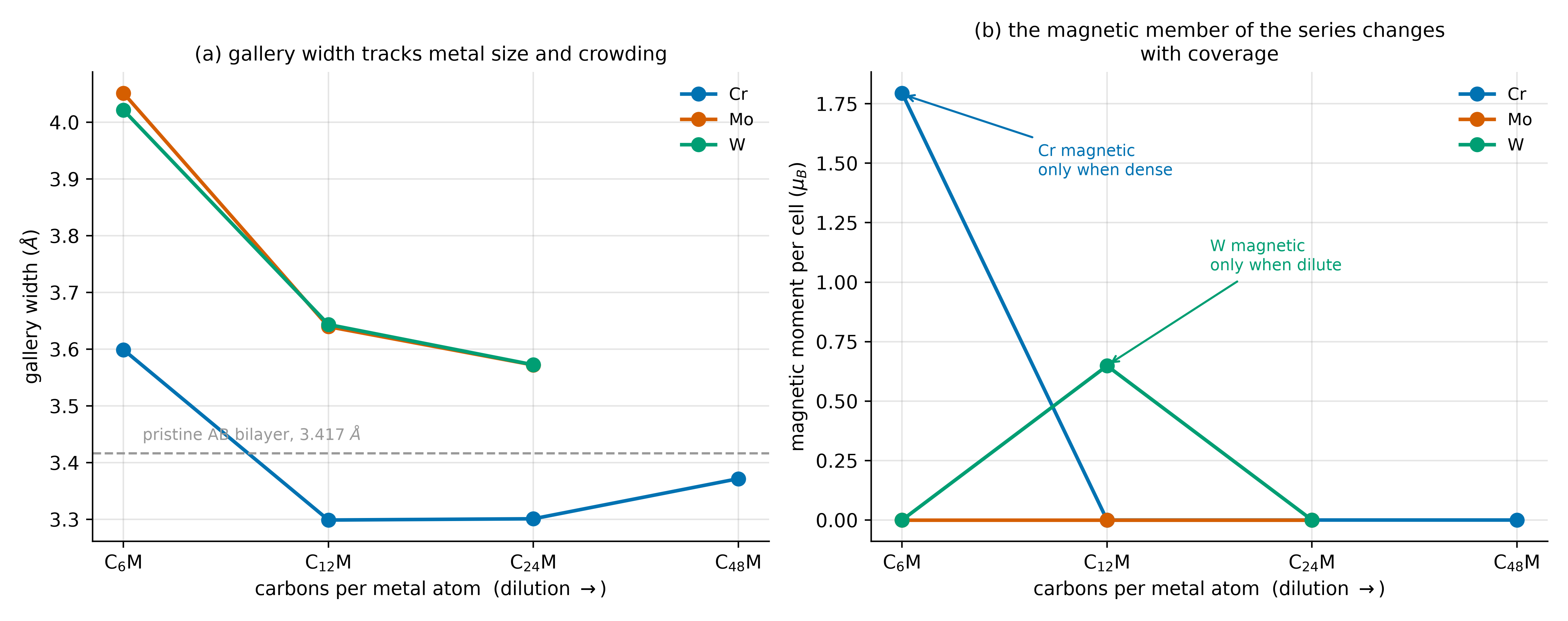}}
\caption*{Figure 7: Coverage dependence across Cr, Mo and W. (a) Gallery
width versus coverage; crowding prises Mo/W sheets apart at C6M. (b)
Magnetic member of the series inverts with coverage: Cr is magnetic at
C6M but not C12M; W is magnetic at C12M but not C6M.}
\end{figure}

\section{7. Magnetism depends on both metal and coverage, and inverts
between
them}\label{magnetism-depends-on-both-metal-and-coverage-and-inverts-between-them}

Extending both coverages across the series gives a pattern that neither
metal identity nor electron counting explains on its own:

\begin{longtable}[]{@{}
  >{\raggedright\arraybackslash}p{(\linewidth - 6\tabcolsep) * \real{0.2500}}
  >{\raggedright\arraybackslash}p{(\linewidth - 6\tabcolsep) * \real{0.2500}}
  >{\raggedright\arraybackslash}p{(\linewidth - 6\tabcolsep) * \real{0.2500}}
  >{\raggedright\arraybackslash}p{(\linewidth - 6\tabcolsep) * \real{0.2500}}@{}}
\caption{Magnetic ground states across coverage and metal for the
group-6 series.}\tabularnewline
\toprule\noalign{}
\begin{minipage}[b]{\linewidth}\raggedright
coverage
\end{minipage} & \begin{minipage}[b]{\linewidth}\raggedright
Cr
\end{minipage} & \begin{minipage}[b]{\linewidth}\raggedright
Mo
\end{minipage} & \begin{minipage}[b]{\linewidth}\raggedright
W
\end{minipage} \\
\midrule\noalign{}
\endfirsthead
\toprule\noalign{}
\begin{minipage}[b]{\linewidth}\raggedright
coverage
\end{minipage} & \begin{minipage}[b]{\linewidth}\raggedright
Cr
\end{minipage} & \begin{minipage}[b]{\linewidth}\raggedright
Mo
\end{minipage} & \begin{minipage}[b]{\linewidth}\raggedright
W
\end{minipage} \\
\midrule\noalign{}
\endhead
\bottomrule\noalign{}
\endlastfoot
C6M (two metals per gallery, metal-metal bonded) & 1.79 muB, FM 81 meV
below NM & 0.00, -1.1 meV & 0.00, +0.5 meV \\
C12M (one metal per gallery, isolated) & 0.00 & 0.00 & 0.65 muB, FM 49
meV below NM \\
\end{longtable}

The magnetic member of the series changes with coverage: chromium at the
dense coverage, tungsten at the dilute one, and in each case the other
two metals admit no magnetic solution at all. The energy differences for
the non-magnetic cases (-1.1 and +0.5 meV) are within numerical noise.

The gallery widths account for part of this:

\begin{longtable}[]{@{}llll@{}}
\caption{Gallery widths at C6M versus C12M coverage.}\tabularnewline
\toprule\noalign{}
system & C6M gallery & C12M gallery & expansion \\
\midrule\noalign{}
\endfirsthead
\toprule\noalign{}
system & C6M gallery & C12M gallery & expansion \\
\midrule\noalign{}
\endhead
\bottomrule\noalign{}
\endlastfoot
Cr & 3.599 A & 3.299 A & +0.30 A \\
Mo & 4.045 A & 3.640 A & +0.41 A \\
W & 4.021 A & 3.634 A & +0.39 A \\
\end{longtable}

At the dense coverage the metals form a bonded honeycomb sublattice
rather than isolated sandwich units, and molybdenum and tungsten --
substantially larger than chromium -- force the sheets
\textasciitilde0.4 A further apart than they do when isolated. The
resulting metal-metal bonding quenches their moments. Chromium, small
enough to occupy the dense lattice without such expansion, retains 1.79
muB; consistent with this, its relaxed Cr-Cr separation of 2.454 A is
\emph{shorter} than the 2.498 A nearest-neighbour distance in bcc
chromium metal.

The 18-electron closed-shell description is therefore not a single rule
governing the series. It applies to isolated metal centres, where it
holds for chromium and molybdenum and fails for tungsten (Section 6); at
dense coverage it is displaced altogether by direct metal-metal bonding,
which is what determines the magnetic behaviour there.

\section{8. What bis-hexahapto intercalation
delivers}\label{what-bis-hexahapto-intercalation-delivers}

The two mechanisms discussed in the Introduction are mutually exclusive
in the relevant electronic respect. Superconductivity in the
intercalated graphites is tied to carbon-derived, interstitial-like
states at E\_F; bis-hexahapto coordination is defined by d-pi
hybridisation that concentrates gallery density on the metal. The
calculations do not support a threshold-style claim that group-6
bilayers ``lack'' interstitial density -- they carry more of it in
absolute terms than the C12Li control -- but they do show consistently
that the fraction of gallery weight surviving atom masking, and the
orbital composition at E\_F, differ from the ionic references in the
direction expected for metal-bound rather than free-electron-like
states.

One inference this comparison does not license should be stated plainly,
because the framing invites it. Metal-d character at the Fermi level
does not by itself preclude phonon-mediated superconductivity: Nb, NbN
and V3Si are d-band superconductors. Two of the ingredients that enter
the coupling constant are in fact more favourable here than in the ionic
controls -- the group-6 bilayers carry two to three times the projected
spectral weight at E\_F, and the narrow d manifold that produces it is
exactly the kind of feature that raises the density of states. What the
present calculations establish is that the states at E\_F are
metal-derived rather than carbon-derived, and therefore that they are
not the interlayer band whose occupancy tracks superconductivity across
the ionic graphite intercalation compounds. Whether they couple instead
to the stiff in-plane C-C modes that supply the high phonon frequencies
of MgB2 and the intercalated graphites is a different question, and it
is settled only by computing the electron-phonon coupling, which we have
not done.

Independently of electronic character, ordered group-6 intercalation is
metastable against segregation into bulk metal (Section 5).
Bis-hexahapto chemistry nonetheless delivers properties well matched to
what the experimental literature reports {[}9,10{]}: strong directional
interlayer bonding of 1.5-4.1 eV per metal against isolated atoms,
preserved metallicity, a definite AA stacking registry imposed wherever
the metal is present (Section 1), and -- for tungsten -- a weak
itinerant moment not predicted by an 18-electron closed-shell count
(Section 6). These are strengths of the chemistry as interlayer
interconnect, not indicators that the materials reproduce the electronic
character of a superconducting graphite intercalation compound.

\section{Conclusions}\label{conclusions}

We asked whether ordered bis-hexahapto group-6 sandwiches between
graphene sheets can support the carbon-derived interlayer-band character
that underlies superconductivity in ionic graphite intercalation
compounds. Density-functional calculations on Cr, Mo and W at C12M and
C6M coverages answer that question in the negative for the orbital
composition at the Fermi level, while clarifying the structural and
thermodynamic conditions under which these sandwiches form.

Three results organise the picture. First, eta6-eta6 coordination
requires AA registry and forbids it in AB; intercalation reverses the
pristine Bernal preference (56 meV per cell) by 400-600 meV, so AA
stacking wherever the metal is present is a sharp, testable prediction.
Second, formation energies are positive against bulk metal (2.9-5.0
eV/atom) but negative against isolated atoms (1.6-4.1 eV), so the
ordered phases are kinetically trapped relative to clustering --
consistent with the experimental Cr \textgreater{} Mo \textgreater{} W
reactivity trend. Third, every group-6 bilayer studied here is metallic,
yet the Fermi-level weight is metal dominated (about 73\% by
element-resolved PDOS, and 89-98\% of that metal weight is d), against a
lower bound of 17\% for bulk CaC6, with atom-masked interstitial
fractions R = 0.15-0.23 against 0.45-0.68 for the matched ionic controls
C12Li and C12Ca. Absolute interstitial weight is not scarce; what
differs is orbital parentage. C12Cr and C12Mo are non-magnetic metals;
an apparent gap in C12Cr appears only when the Brillouin-zone mesh
misses K. Tungsten alone carries a weak ferromagnetic moment of 0.65
muB, so an 18-electron closed-shell count does not describe the full
group-6 series.

Bis-hexahapto intercalation therefore remains well matched to
directional interlayer bonding and electrical interconnection of carbon
surfaces, but it does not reproduce the carbon-dominated Fermi-level
character of superconducting ionic GICs. That is a statement about which
states carry the Fermi surface, not a prediction that these materials
cannot superconduct: d-band superconductors exist, and the group-6
bilayers carry more spectral weight at E\_F than either ionic control.
Deciding the question requires the electron-phonon coupling, not the
orbital character.

Zone-centre phonons place no lattice instability in any of the three
ordered C12M bilayers, with the numerical noise floor 2.6 to 16 times
below the softest optical branch; the pristine bilayer computed as a
control in both stackings separates as it should -- Bernal AB is stable,
AA carries an imaginary interlayer shear at -29 cm-1 against +22 cm-1 in
AB -- so the metal is what holds the registry. That is a zone-centre
statement only: a q = 0 calculation cannot establish dynamical
stability, and the C6Cr bond alternation reported above is a
zone-boundary effect invisible to it. A full q mesh, migration barriers,
electron-phonon coupling and a fully self-consistent relativistic
treatment of tungsten were not computed and remain the natural next
steps.

\section{References}\label{references}

{[}1{]} T. E. Weller, M. Ellerby, S. S. Saxena, R. P. Smith and N. T.
Skipper, ``Superconductivity in the intercalated graphite compounds C6Yb
and C6Ca'', \emph{Nature Physics} 1, 39-41 (2005).

{[}2{]} M. Calandra and F. Mauri, ``Theoretical explanation of
superconductivity in C6Ca'', \emph{Physical Review Letters} 95, 237002
(2005). DOI 10.1103/PhysRevLett.95.237002.

{[}3{]} G. Csányi, P. B. Littlewood, A. H. Nevidomskyy, C. J. Pickard
and B. D. Simons, ``The role of the interlayer state in the electronic
structure of superconducting graphite intercalated compounds'',
\emph{Nature Physics} 1, 42-45 (2005).

{[}4{]} G. Profeta, M. Calandra and F. Mauri, ``Phonon-mediated
superconductivity in graphene by lithium deposition'', \emph{Nature
Physics} 8, 131-134 (2012). DOI 10.1038/nphys2181.

{[}5{]} B. M. Ludbrook \emph{et al.}, ``Evidence for superconductivity
in Li-decorated monolayer graphene'', \emph{Proceedings of the National
Academy of Sciences} 112, 11795-11799 (2015). DOI
10.1073/pnas.1510435112.

{[}6{]} S. Ichinokura, K. Sugawara, A. Takayama, T. Takahashi and S.
Hasegawa, ``Superconducting calcium-intercalated bilayer graphene'',
\emph{ACS Nano} 10, 2761-2765 (2016). DOI 10.1021/acsnano.5b07848.

{[}7{]} S. Sarkar, S. Niyogi, E. Bekyarova and R. C. Haddon,
``Organometallic chemistry of extended periodic pi-electron systems:
hexahapto-chromium complexes of graphene and single-walled carbon
nanotubes'', \emph{Chemical Science} 2, 1326-1333 (2011). DOI
10.1039/C0SC00634C.

{[}8{]} A. Pekker, M. Chen, E. Bekyarova and R. C. Haddon,
``Photochemical generation of bis-hexahapto chromium interconnects
between the graphene surfaces of single-walled carbon nanotubes'',
\emph{Materials Horizons} 2, 81-85 (2015). DOI 10.1039/C4MH00192C.

{[}9{]} M. Chen, X. Tian, W. Li, E. Bekyarova, G. Li, M. Moser and R. C.
Haddon, ``Application of organometallic chemistry to the electrical
interconnection of graphene nanoplatelets'', \emph{Chemistry of
Materials} 28, 2260-2266 (2016). DOI 10.1021/acs.chemmater.6b00217.

{[}10{]} M. Chen, \emph{Chemical Engineering of Graphene and
Single-Walled Carbon Nanotubes for Electronic Applications},
Ph.D.~dissertation, University of California, Riverside, September 2017.
Permalink https://escholarship.org/uc/item/1fm925nh.

{[}11{]} A. V. Krasheninnikov, P. O. Lehtinen, A. S. Foster, P. Pyykkö
and R. M. Nieminen, ``Embedding transition-metal atoms in graphene:
structure, bonding, and magnetism'', \emph{Physical Review Letters} 102,
126807 (2009).

{[}12{]} K. P. Lucht, A. D. Mahabir, A. Alcantara, A. V. Balatsky, J. L.
Mendoza-Cortes and J. T. Haraldsen, ``Designing a path towards
superconductivity through magnetic exchange in transition-metal
intercalated bilayer graphene'', arXiv:1903.10112 (2019).

{[}13{]} D. Li, Z. Gui, M. Ling, L. Guo, Z. Wang, Q. Yuan and L. Cheng,
``Modulating the bandgap of Cr-intercalated bilayer graphene via
combining the 18-electron rule and the 2D superatomic-molecule theory'',
\emph{Nanoscale} 16, 17433-17441 (2024). DOI 10.1039/D4NR02440K.

{[}14{]} J. Enkovaara \emph{et al.}, ``Electronic structure calculations
with GPAW: a real-space implementation of the projector augmented-wave
method'', \emph{Journal of Physics: Condensed Matter} 22, 253202 (2010).
DOI 10.1088/0953-8984/22/25/253202.

{[}15{]} J. J. Mortensen, L. B. Hansen and K. W. Jacobsen, ``Real-space
grid implementation of the projector augmented wave method'',
\emph{Physical Review B} 71, 035109 (2005).

{[}16{]} A. Hjorth Larsen \emph{et al.}, ``The atomic simulation
environment -- a Python library for working with atoms'', \emph{Journal
of Physics: Condensed Matter} 29, 273002 (2017). DOI
10.1088/1361-648X/aa680e.

{[}17{]} J. P. Perdew, K. Burke and M. Ernzerhof, ``Generalized gradient
approximation made simple'', \emph{Physical Review Letters} 77,
3865-3868 (1996); erratum \emph{ibid.} 78, 1396 (1997).

{[}18{]} S. Grimme, J. Antony, S. Ehrlich and H. Krieg, ``A consistent
and accurate ab initio parametrization of density functional dispersion
correction (DFT-D) for the 94 elements H-Pu'', \emph{The Journal of
Chemical Physics} 132, 154104 (2010). DOI 10.1063/1.3382344.

{[}19{]} S. Grimme, S. Ehrlich and L. Goerigk, ``Effect of the damping
function in dispersion corrected density functional theory'',
\emph{Journal of Computational Chemistry} 32, 1456-1465 (2011).

\appendix
\renewcommand{\thetable}{S\arabic{table}}
\setcounter{table}{0}
\section{Supporting Information}\label{supporting-information}

\subsection{1. SCF setting sweeps on
C12Cr}\label{scf-setting-sweeps-on-c12cr}

All tests on C12Cr in the sqrt(3)xsqrt(3) cell, varying one setting at a
time from the adopted baseline (plane waves, 450 eV, 9x9x1, Fermi-Dirac
0.05 eV, 9 A vacuum, standard Cr PAW dataset, non-spin-polarised unless
stated).

\begin{quote}
The \texttt{gap\ proxy} column is not a band gap. It is
\texttt{max(E\ below\ E\_F)\ -\textgreater{}\ min(E\ at\ or\ above\ E\_F)}
measured on the SCF integration mesh. For a system whose bands cross
E\_F this measures the spacing between discrete eigenvalues straddling
E\_F, which depends on the mesh and on where the smearing places E\_F
rather than on the material. It is tabulated here to document that
artefact. Band gaps reported in the main text are determined by band
index on a dense uniform mesh (see Methods).
\end{quote}

\subsection{Plane-wave cutoff}\label{plane-wave-cutoff}

{\def\LTcaptype{none} % do not increment counter
\begin{longtable}[]{@{}lllll@{}}
\toprule\noalign{}
setting & IBZ k-points & E / eV & gap proxy / eV & wall / s \\
\midrule\noalign{}
\endhead
\bottomrule\noalign{}
\endlastfoot
350 & 25 & -116.1011 & 0.1126 & 95 \\
400 & 25 & -117.8876 & 0.1088 & 100 \\
450 & 25 & -118.5820 & 0.1112 & 118 \\
550 & 25 & -118.8802 & 0.1116 & 217 \\
650 & 25 & -118.9163 & 0.1114 & 240 \\
\end{longtable}
}

Spread in the proxy across cutoffs: 3.8 meV. Absolute energies are not
comparable between different cutoffs (basis-set incompleteness shifts
the reference), so only the proxy is compared here.

\subsection{k-point mesh}\label{k-point-mesh}

{\def\LTcaptype{none} % do not increment counter
\begin{longtable}[]{@{}lllll@{}}
\toprule\noalign{}
setting & IBZ k-points & E / eV & gap proxy / eV & wall / s \\
\midrule\noalign{}
\endhead
\bottomrule\noalign{}
\endlastfoot
6x6x1 & 18 & -118.5875 & 0.2668 & 89 \\
9x9x1 & 25 & -118.5820 & 0.1112 & 118 \\
15x15x1 & 64 & -118.5845 & 0.0325 & 339 \\
18x18x1 & 162 & -118.5844 & 0.0138 & 791 \\
\end{longtable}
}

Total energy varies by 5.5 meV over this range while the proxy varies by
253 meV (a factor of 46). An energy-based convergence test therefore
certifies a mesh as converged while a gap read off that same mesh is
still wrong by an order of magnitude.

\subsection{Smearing width}\label{smearing-width}

{\def\LTcaptype{none} % do not increment counter
\begin{longtable}[]{@{}lllll@{}}
\toprule\noalign{}
setting & IBZ k-points & E / eV & gap proxy / eV & wall / s \\
\midrule\noalign{}
\endhead
\bottomrule\noalign{}
\endlastfoot
0.02 & 25 & -118.5728 & 0.0056 & 138 \\
0.05 & 25 & -118.5820 & 0.1112 & 118 \\
0.1 & 25 & -118.5999 & 0.1047 & 123 \\
\end{longtable}
}

The proxy is non-monotonic in the smearing width (0.02 eV
-\textgreater{} 0.0056 eV, 0.05 eV -\textgreater{} 0.1112 eV, 0.1 eV
-\textgreater{} 0.1047 eV), which a genuine gap cannot be. This is the
signature of E\_F moving between discrete levels and is the clearest
single diagnostic that the proxy is not measuring a gap.

\subsection{Vacuum thickness}\label{vacuum-thickness}

{\def\LTcaptype{none} % do not increment counter
\begin{longtable}[]{@{}lllll@{}}
\toprule\noalign{}
setting & IBZ k-points & E / eV & gap proxy / eV & wall / s \\
\midrule\noalign{}
\endhead
\bottomrule\noalign{}
\endlastfoot
7 & 25 & -118.5816 & 0.1118 & 98 \\
9 & 25 & -118.5820 & 0.1112 & 118 \\
12 & 25 & -118.5813 & 0.1118 & 143 \\
\end{longtable}
}

\subsection{PAW valence}\label{paw-valence}

{\def\LTcaptype{none} % do not increment counter
\begin{longtable}[]{@{}lllll@{}}
\toprule\noalign{}
setting & IBZ k-points & E / eV & gap proxy / eV & wall / s \\
\midrule\noalign{}
\endhead
\bottomrule\noalign{}
\endlastfoot
standard & 25 & -118.5820 & 0.1112 & 118 \\
Cr:14 & 25 & -118.5471 & 0.1138 & 140 \\
\end{longtable}
}

\subsection{Spin polarisation}\label{spin-polarisation}

{\def\LTcaptype{none} % do not increment counter
\begin{longtable}[]{@{}lllll@{}}
\toprule\noalign{}
setting & IBZ k-points & E / eV & gap proxy / eV & wall / s \\
\midrule\noalign{}
\endhead
\bottomrule\noalign{}
\endlastfoot
no & 25 & -118.5820 & 0.1112 & 118 \\
spin\_check & -- & error & -- & -- \\
\end{longtable}
}

\subsection{2. Version-2 interlayer metric: mesh and parameter
sensitivity}\label{version-2-interlayer-metric-mesh-and-parameter-sensitivity}

Primary observable on production-refined \texttt{\_tight} geometries: R
= sum(q\_i I\_i) / sum(q\_i G\_i), the k- and occupation-weighted
fraction of gallery spectral weight that survives masking of all atomic
spheres (radius 1.30 A, gallery fraction 0.45, Gaussian kernel sigma =
0.10 eV). Absolute interstitial weight A = sum(q\_i I\_i) and raw
gallery weight G = sum(q\_i G\_i) are reported separately. Generated
from \texttt{interlayer\_convergence.json}.

\subsection{Mesh convergence of R}\label{mesh-convergence-of-r}

{\def\LTcaptype{none} % do not increment counter
\begin{longtable}[]{@{}lllllll@{}}
\toprule\noalign{}
system & R (12x12) & R (18x18) & & R12-R18 & & A (12x12) \\
\midrule\noalign{}
\endhead
\bottomrule\noalign{}
\endlastfoot
bulk CaC6 & 0.698 & 0.698 & 0.0004 & 0.0893 & 0.1281 & yes \\
C12Ca & 0.682 & 0.695 & 0.0127 & 0.2285 & 0.3349 & yes \\
C12Li & 0.448 & 0.446 & 0.0023 & 0.0629 & 0.1404 & yes \\
C12Cr & 0.152 & 0.152 & 0.0001 & 0.0860 & 0.5667 & yes \\
C12Mo & 0.223 & 0.223 & 0.0002 & 0.1373 & 0.6151 & yes \\
C12W & 0.229 & 0.230 & 0.0005 & 0.1216 & 0.5308 & yes \\
\end{longtable}
}

All systems pass the 12x12 -\textgreater{} 18x18 drift tolerance
relative to the baseline R. Mesh drift is largest for C12Ca (0.013) and
negligible for the group-6 bilayers (\textless0.001).

\subsection{Non-mesh parameter
envelopes}\label{non-mesh-parameter-envelopes}

Each envelope spans the baseline setting plus variations of mask radius
(1.10 / 1.30 / 1.50 A), gallery fraction (0.35 / 0.45 / 0.55), kernel
width (0.05 / 0.10 / 0.15 eV), and an occupation-derivative kernel.
Envelopes are ranges over numerical settings, not statistical
uncertainties; they must not be plotted as error bars.

{\def\LTcaptype{none} % do not increment counter
\begin{longtable}[]{@{}llll@{}}
\toprule\noalign{}
system & R baseline & envelope min & envelope max \\
\midrule\noalign{}
\endhead
\bottomrule\noalign{}
\endlastfoot
bulk CaC6 & 0.698 & 0.534 & 0.800 \\
C12Ca & 0.682 & 0.502 & 0.794 \\
C12Li & 0.448 & 0.227 & 0.805 \\
C12Cr & 0.152 & 0.053 & 0.336 \\
C12Mo & 0.223 & 0.088 & 0.457 \\
C12W & 0.229 & 0.095 & 0.464 \\
\end{longtable}
}

\subsection{Reading the envelopes}\label{reading-the-envelopes}

Baseline R(C12Ca) - R(C12Li) = 0.234 (condition that the ionic same-cell
controls separate: pass). The group-6 envelopes overlap the C12Li
envelope when each system's \emph{worst} setting is compared against the
control's \emph{best} setting; Li\_min - group6\_max margins are Cr =
-0.109, Mo = -0.230, W = -0.237. That overlap is almost entirely
mask-radius driven: C12Li's envelope minimum occurs at radius 1.50 A,
while the group-6 maxima occur at radius 1.10 A. A post-hoc,
matched-parameter check (same radius, gallery fraction and kernel for
both systems) finds R(group-6) \textless{} R(C12Li) at every common
setting; that observation is legitimate for the SI but does not
reinstate a threshold-style exclusion claim in the main text. The paper
reports R and A descriptively, with envelopes shown as envelopes.

\subsection{3. Element composition at E\_F from projected
DOS}\label{element-composition-at-e_f-from-projected-dos}

Element- and l-resolved PDOS from \texttt{tmbg\_bands.projected\_dos},
integrated in \textbar E - E\_F\textbar{} \(\leq\) 0.15 eV and
normalised by element over the bound s/p/d projector channels each PAW
dataset provides. Generated from \texttt{orbital\_character.json}.

\begin{quote}
\emph{Why not metal }d* against carbon \emph{p}.* GPAW builds a
projected density of states from bound PAW projectors only
(\texttt{gpaw.dos.get\_projector\_numbers},
\texttt{n\ \textgreater{}=\ 0}). In the default datasets the Ca \emph{d}
projector is unbound and Li has no \emph{d} projector at all, so a
Ca-\emph{d} or Li-\emph{d} PDOS is identically zero over the whole
energy range as a property of the basis, not of the material, while
Cr/Mo/W report real bound-\emph{d} weight. Normalising by element
removes that asymmetry. The residual limitation is that a
bound-projector scheme cannot register Ca-\emph{d} hybridisation, so the
metal fractions quoted for Ca and Li are lower bounds.
\end{quote}

{\def\LTcaptype{none} % do not increment counter
\begin{longtable}[]{@{}
  >{\raggedright\arraybackslash}p{(\linewidth - 10\tabcolsep) * \real{0.1667}}
  >{\raggedright\arraybackslash}p{(\linewidth - 10\tabcolsep) * \real{0.1667}}
  >{\raggedright\arraybackslash}p{(\linewidth - 10\tabcolsep) * \real{0.1667}}
  >{\raggedright\arraybackslash}p{(\linewidth - 10\tabcolsep) * \real{0.1667}}
  >{\raggedright\arraybackslash}p{(\linewidth - 10\tabcolsep) * \real{0.1667}}
  >{\raggedright\arraybackslash}p{(\linewidth - 10\tabcolsep) * \real{0.1667}}@{}}
\toprule\noalign{}
\begin{minipage}[b]{\linewidth}\raggedright
system
\end{minipage} & \begin{minipage}[b]{\linewidth}\raggedright
metal (\%)
\end{minipage} & \begin{minipage}[b]{\linewidth}\raggedright
carbon (\%)
\end{minipage} & \begin{minipage}[b]{\linewidth}\raggedright
\emph{d} as \% of metal
\end{minipage} & \begin{minipage}[b]{\linewidth}\raggedright
bound channels
\end{minipage} & \begin{minipage}[b]{\linewidth}\raggedright
classification
\end{minipage} \\
\midrule\noalign{}
\endhead
\bottomrule\noalign{}
\endlastfoot
bulk CaC6 & \textgreater{} 17 & 83 & n/a & Ca: s/p & carbon dominated \\
C12Ca & \textgreater{} 2 & 98 & n/a & Ca: s/p & carbon dominated \\
C12Li & \textgreater{} 13 & 87 & n/a & Li: s/p & carbon dominated \\
C12Cr & 72 & 28 & 98 & Cr: s/p/d & metal dominated \\
C12Mo & 74 & 26 & 91 & Mo: s/p/d & metal dominated \\
C12W & 73 & 27 & 89 & W: s/p/d & metal dominated \\
\end{longtable}
}

Ionic controls and bulk CaC6 are carbon dominated (yes); all three
group-6 bilayers are metal dominated (yes), over a range of 72-74 per
cent against a largest control lower bound of 17 per cent.

The published v1 of this table normalised over metal \emph{d} plus
carbon \emph{p} and reported the ionic controls as 0 per cent metal
\emph{d}. Those values are retained in \texttt{orbital\_character.json}
under \texttt{legacy\_two\_channel\_*} for traceability and are
superseded here.

\subsection{4. CaC6 interlayer band identified from
dispersion}\label{cac6-interlayer-band-identified-from-dispersion}

\texttt{ANALYSIS\_SPEC.md} condition 6 asks that the bulk CaC6 benchmark
identify an interstitial state at E\_F whose character is confirmed
directly. \texttt{CLAIM\_AUDIT.md} Section 3a records why the
planar-average attempt failed: the intercalant sits at the gallery
midplane in CaC6, so ``between the sheets'' and ``on the metal'' are the
same z. This test uses eigenvalues only -- no atom mask and no PAW
projection, so it cannot inherit the assumptions of the measure it
validates.

A nearly free-electron interlayer state is delocalised in the gallery
and disperses in three dimensions: a parabolic minimum at the zone
centre with a light, nearly isotropic mass. A carbon \(\pi\) band in a
layered crystal is light in plane and heavy across the layers. Generated
from \texttt{interlayer\_band\_validity.json}.

{\def\LTcaptype{none} % do not increment counter
\begin{longtable}[]{@{}
  >{\raggedright\arraybackslash}p{(\linewidth - 8\tabcolsep) * \real{0.2000}}
  >{\raggedright\arraybackslash}p{(\linewidth - 8\tabcolsep) * \real{0.2000}}
  >{\raggedright\arraybackslash}p{(\linewidth - 8\tabcolsep) * \real{0.2000}}
  >{\raggedright\arraybackslash}p{(\linewidth - 8\tabcolsep) * \real{0.2000}}
  >{\raggedright\arraybackslash}p{(\linewidth - 8\tabcolsep) * \real{0.2000}}@{}}
\toprule\noalign{}
\begin{minipage}[b]{\linewidth}\raggedright
band
\end{minipage} & \begin{minipage}[b]{\linewidth}\raggedright
E at \(\Gamma\) (eV)
\end{minipage} & \begin{minipage}[b]{\linewidth}\raggedright
m*/m\_e in-plane
\end{minipage} & \begin{minipage}[b]{\linewidth}\raggedright
m*/m\_e out-of-plane
\end{minipage} & \begin{minipage}[b]{\linewidth}\raggedright
anisotropy
\end{minipage} \\
\midrule\noalign{}
\endhead
\bottomrule\noalign{}
\endlastfoot
54 & -1.266 & 0.49 & 0.47 & 1.03 \\
52 & -1.882 & 0.20 & 0.79 & 3.88 \\
53 & -1.878 & 0.20 & 0.79 & 3.94 \\
\end{longtable}
}

Every band listed crosses E\_F and has its minimum at \(\Gamma\). Band
54 is the only isotropic one (anisotropy 1.03 against 3.9 and above for
the carbon bands), with a light mass of 0.49 m\_e and a minimum 1.27 eV
below E\_F. That is the interlayer band, partially occupied, reproducing
the qualitative criterion of Csányi and co-workers. Condition 6 is met:
identified.

The mask-based rule selects a different band, and that is expected.
\texttt{construct\_validity.json} picks the single near-E\_F state with
the largest masked interstitial fraction, and lands on band 51 (at
+0.005 eV), not band 54. The two rules optimise different quantities --
largest single-state interstitial fraction against free-electron
dispersion -- so they need not agree, and Section 3a of the claim audit
already records that the per-state selection is degenerate. The
disagreement is about which band to name, not about the physics:
averaged over states within 0.15 eV of E\_F, both carry high
interstitial character -- band 51 at I/G = 0.78, band 54 at I/G = 0.54,
against 0.15-0.23 for the group-6 bilayers. Band indices are directly
comparable between the two records: both calculations store the same 66
bands in the same order.

\subsection{5. Tungsten spin-orbit coupling
check}\label{tungsten-spin-orbit-coupling-check}

Non-self-consistent SOC eigenvalues on the production-refined
\texttt{C12M\_W\_AA\_FM\_tight} density (GPAW soc\_eigenstates,
non-self-consistent fixed density; mesh 12x12x1; smearing 0.02 eV).

{\def\LTcaptype{none} % do not increment counter
\begin{longtable}[]{@{}
  >{\raggedright\arraybackslash}p{(\linewidth - 2\tabcolsep) * \real{0.5000}}
  >{\raggedright\arraybackslash}p{(\linewidth - 2\tabcolsep) * \real{0.5000}}@{}}
\toprule\noalign{}
\begin{minipage}[b]{\linewidth}\raggedright
quantity
\end{minipage} & \begin{minipage}[b]{\linewidth}\raggedright
value
\end{minipage} \\
\midrule\noalign{}
\endhead
\bottomrule\noalign{}
\endlastfoot
magnetic anisotropy E(x) - E(z) & 2.11 meV \\
SOC band-energy correction & -0.173 eV \\
metallic under SOC & yes (gap = 0.000 eV; 80 partially occupied
states) \\
\end{longtable}
}

SOC does not open a gap and does not overturn the metallic character
reported for C12W in the main text. The check is non-self-consistent; a
fully relativistic treatment remains desirable for a complete
characterisation of the tungsten ground state.

\subsection{6. Zone-centre phonons}\label{zone-centre-phonons}

Finite displacements (delta = 0.01 Angstrom, both signs, all 3N
Cartesian degrees of freedom) on the production \texttt{\_tight}
geometries, with the production calculator and D3 contributing to the
force constants. The acceptance criterion was frozen in
\texttt{ANALYSIS\_SPEC.md} section 8 before any frequency was computed.

\begin{quote}
\emph{Scope.} The supercell is (1,1,1), so this resolves q = 0 of the
sqrt(3)xsqrt(3)R30 cell. That is more than the zone centre of graphene:
the sqrt(3) cell folds the primitive-cell K point onto Gamma, so these
spectra also contain the carbon K-point modes. What is NOT probed is the
primitive-cell M point, and the zone boundary of the metal superlattice
-- for which Gamma of this cell is only the zone centre. The
non-magnetic C6Cr bond alternation reported in the main text is a
metal-sublattice zone-boundary instability in a different cell again,
and remains invisible here by construction. These calculations can
exclude a zone-centre instability; they cannot establish dynamical
stability, and no such claim is made. A full q mesh needs a 2x2
supercell -- 52 atoms and 312 displacements -- which was out of reach on
the machine used.
\end{quote}

\texttt{residual} is the largest \textbar omega\textbar{} among the
three acoustic branches with the acoustic sum rule off: the three
translations vanish in an exact calculation, so it measures the noise
floor of the force constants, not a physical frequency.
\texttt{post-ASR} is the same quantity after the sum rule, reported in
addition because the rule is applied iteratively with symmetrisation and
does not always drive the translations closer to zero. Rule 2 decides on
\texttt{residual}; the margin column uses the larger of the two, which
is the stricter reading.

{\def\LTcaptype{none} % do not increment counter
\begin{longtable}[]{@{}
  >{\raggedright\arraybackslash}p{(\linewidth - 12\tabcolsep) * \real{0.1429}}
  >{\raggedright\arraybackslash}p{(\linewidth - 12\tabcolsep) * \real{0.1429}}
  >{\raggedright\arraybackslash}p{(\linewidth - 12\tabcolsep) * \real{0.1429}}
  >{\raggedright\arraybackslash}p{(\linewidth - 12\tabcolsep) * \real{0.1429}}
  >{\raggedright\arraybackslash}p{(\linewidth - 12\tabcolsep) * \real{0.1429}}
  >{\raggedright\arraybackslash}p{(\linewidth - 12\tabcolsep) * \real{0.1429}}
  >{\raggedright\arraybackslash}p{(\linewidth - 12\tabcolsep) * \real{0.1429}}@{}}
\toprule\noalign{}
\begin{minipage}[b]{\linewidth}\raggedright
system
\end{minipage} & \begin{minipage}[b]{\linewidth}\raggedright
residual (cm\^{}-1)
\end{minipage} & \begin{minipage}[b]{\linewidth}\raggedright
post-ASR
\end{minipage} & \begin{minipage}[b]{\linewidth}\raggedright
softest optical
\end{minipage} & \begin{minipage}[b]{\linewidth}\raggedright
margin
\end{minipage} & \begin{minipage}[b]{\linewidth}\raggedright
imaginary
\end{minipage} & \begin{minipage}[b]{\linewidth}\raggedright
verdict
\end{minipage} \\
\midrule\noalign{}
\endhead
\bottomrule\noalign{}
\endlastfoot
C12Cr & 9.7 & 10.2 & 163.4 & 16.1x & 0 & no zone centre instability \\
C12Mo & 29.4 & 42.7 & 112.3 & 2.6x & 0 & no zone centre instability \\
C12W & 18.4 & 15.5 & 151.6 & 8.2x & 0 & no zone centre instability \\
pristine bilayer, AA † & 18.8 & 15.8 & 28.9 & 1.5x & 2 & zone centre
instability \\
pristine bilayer, AB † & 16.0 & 15.6 & 21.8 & 1.4x & 0 & no zone centre
instability \\
\end{longtable}
}

† Rows marked with a dagger (pristine bilayer, AA; pristine bilayer, AB)
clear their own noise floor by less than 2x. The verdict column still
follows spec rule 2, which these rows pass, but at that margin only the
SIGN of the softest branch is meaningful, not its magnitude. The
AA-vs-AB comparison in the main text rests on the sign, and on a
\textasciitilde50 cm\^{}-1 difference that is far outside either noise
floor.

Imaginary modes, with the displacement pattern spec section 8.4 rule 4
requires. \texttt{antiphase} is the cosine between the two carbon
sheets' mean in-plane displacement: near -1 the sheets slide against
each other, which is the interlayer shear coordinate.

{\def\LTcaptype{none} % do not increment counter
\begin{longtable}[]{@{}
  >{\raggedright\arraybackslash}p{(\linewidth - 10\tabcolsep) * \real{0.1667}}
  >{\raggedright\arraybackslash}p{(\linewidth - 10\tabcolsep) * \real{0.1667}}
  >{\raggedright\arraybackslash}p{(\linewidth - 10\tabcolsep) * \real{0.1667}}
  >{\raggedright\arraybackslash}p{(\linewidth - 10\tabcolsep) * \real{0.1667}}
  >{\raggedright\arraybackslash}p{(\linewidth - 10\tabcolsep) * \real{0.1667}}
  >{\raggedright\arraybackslash}p{(\linewidth - 10\tabcolsep) * \real{0.1667}}@{}}
\toprule\noalign{}
\begin{minipage}[b]{\linewidth}\raggedright
system
\end{minipage} & \begin{minipage}[b]{\linewidth}\raggedright
omega (cm\^{}-1)
\end{minipage} & \begin{minipage}[b]{\linewidth}\raggedright
above noise
\end{minipage} & \begin{minipage}[b]{\linewidth}\raggedright
in-plane fraction
\end{minipage} & \begin{minipage}[b]{\linewidth}\raggedright
antiphase
\end{minipage} & \begin{minipage}[b]{\linewidth}\raggedright
character
\end{minipage} \\
\midrule\noalign{}
\endhead
\bottomrule\noalign{}
\endlastfoot
pristine bilayer, AA & -158.9 & 8.44x & 0.119 & -0.730 & out-of-plane
(breathing / buckling) \\
pristine bilayer, AA & -28.9 & 1.54x & 1.000 & -1.000 & interlayer shear
(sheets slide against each other) \\
\end{longtable}
}

Graphene's in-plane C-C stretch is measured near 1580 cm\^{}-1, and the
highest branch of every system reproduces it: C12Cr 1583; C12Mo 1603;
C12W 1606; pristine bilayer, AA 1593; pristine bilayer, AB 1577
cm\^{}-1, all within 1.7 per cent. Nothing was fitted to it, so it is an
independent check that the force constants are right.

\end{document}